\documentclass[letterpaper,english,aps,prx,superscriptaddress,twocolumn,amsfonts, amssymb]{revtex4}
\usepackage[T1]{fontenc}
\pdfoutput=1
\usepackage{textcomp}
\usepackage{mathrsfs}
\usepackage{amsmath}
\usepackage{amssymb}
\usepackage{graphicx}
\usepackage{esint}

\makeatletter
\usepackage[english]{babel}

\usepackage[bookmarks=true,colorlinks,linkcolor=blue,urlcolor=blue,citecolor=blue]{hyperref}

\newcommand{\be}{\begin{equation}}
\newcommand{\ee}{\end{equation}}
\newcommand{\bea}{\begin{eqnarray}}
\newcommand{\eea}{\end{eqnarray}}

\newcommand{\mc}{\mathcal}

\usepackage{epsfig,graphicx,psfrag,amsmath,amssymb,float}
\usepackage[caption=false]{subfig}

\@ifundefined{showcaptionsetup}{}{%
 \PassOptionsToPackage{caption=false}{subfig}}
\usepackage{subfig}
\makeatother

\usepackage{babel}

\begin{document}

\title{SU(4)-symmetric  spin-orbital liquids on the hyperhoneycomb
lattice }

\author{Willian M. H. Natori}

\affiliation{Instituto de F\'isica de S\~ao Carlos, Universidade de S\~ao Paulo, CP
369, S\~ao Carlos, SP, 13560-970, Brazil}

\author{Eric C. Andrade }

\affiliation{Instituto de F\'isica de S\~ao Carlos, Universidade de S\~ao Paulo, CP
369, S\~ao Carlos, SP, 13560-970, Brazil}

\author{Rodrigo G. Pereira}
 
\affiliation{International Institute of Physics and Departamento de F\'isica Te\'orica
e Experimental, Universidade Federal do Rio Grande do Norte, 
Natal, RN, 59078-970, Brazil}

\begin{abstract}

We study the effective spin-orbital model that describes the magnetism  of 4$d^1$ or 5$d^1$ Mott insulators in ideal tricoordinated lattices. In the limit of vanishing Hund's coupling, the model has an emergent SU(4) symmetry which is made explicit by means of a Klein transformation on pseudospin degrees of freedom. Taking the hyperhoneycomb lattice as an example, we employ  parton constructions with  fermionic representations of the pseudospin  operators to  investigate possible quantum spin-orbital liquid states. We then use variational Monte Carlo (VMC) methods to compute the energies of the projected wave functions. Our numerical  results show that the lowest-energy quantum liquid   corresponds to a zero-flux state  with a  Fermi surface of four-color fermionic partons. In spite of the Fermi surface, we demonstrate that this state is stable against tetramerization. A combination of linear flavor wave theory and VMC applied to the complete microscopic model also shows that this liquid state is stable against the formation of collinear long-range order.

\end{abstract}
\maketitle

%\section{Introduction}

\section{Introduction}

The search for unconventional phases  induced by  the combined effects of strong correlation and spin-orbit coupling  has stimulated  the study of  transition metal oxides with 4d and 5d elements  \cite{Witczak-Krempa2014,Rau2016,Trebst2017,Winter2017}. Particularly interesting in this context is the demonstration by Jackeli and Khaliullin \cite{Jackeli2009} that the effective spin model for Mott insulators with heavy $d^5$   ions in edge-sharing octahedral geometries contains bond-dependent  Ising-like exchange interactions.  Such interactions constitute the key ingredient of Kitaev's honeycomb  model \cite{Kitaev2006}, an exactly solvable spin-$1/2$ model   with a quantum spin liquid ground state  \cite{Savary2016,Zhou2017}.  Indeed, experiments have shown that Kitaev-type interactions are relevant for the honeycomb  iridates  \cite{SinghPRL2012,Choi2012,Chun2015} and for {$\alpha$-RuCl$_3$}  \cite{PlumbPRB2014,Sears2015,BanerjeeNatMat2016}, in which  the Ir$^{4+}$ or Ru$^{3+}$ ions form $j=1/2$ local moments. In addition, the physics of Kitaev spin liquids has been generalized to tricoordinated three-dimensional lattices \cite{Mandal2009,Hermanns2014,Hermanns2015,OBrien2016}. One example is the hyperhoneycomb lattice, which is materialized  in    $\beta$-Li$_2$IrO$_3$ \cite{BiffinB2014,Takayama2015}. However, the realization of  quantum spin liquids in  the strong spin-orbit coupling regime  has remained a challenge because more realistic models for these compounds  include additional interactions that tend to drive different kinds of long-range magnetic order \cite{Chaloupka2010,Reuther2011,Kimchi2011,Lee2014,Rau2014,Winter2016,Nishimoto2016,Katukuri2016,Koga2018}.  

An alternative recipe for quantum spin liquids may come from substituting the $d^5$  by $d^1$   configuration in the same octahedral environment. In this case, the single electron in the open shell occupies a low-energy  $j=3/2$  quadruplet \cite{Khaliullin2005,khomskii2014transition}. Despite the larger moment, $j=3/2$ systems are not necessarily more classical than their $j=1/2$ counterparts since they can exhibit unexpected continuous symmetries that enhance quantum fluctuations.  For instance, the effective spin model for heavy-element double perovskites with $d^1$ configuration contains bond-dependent interactions with a hidden SU(2) symmetry \cite{Chen2010}. This SU(2)  symmetry  is made explicit when the model is expressed in terms of pseudospin and pseudo-orbital operators    \cite{Natori2016,Romhanyi2017,Natori2017}, and its effects motivated the proposal of a quantum spin-orbital liquid in double perovskites \cite{Natori2016}. Even more surprisingly, it was recently shown  that the spin model for $j=3/2$ moments on several tricoordinated lattices, including the hyperhoneycomb, has an emergent SU(4) symmetry \cite{Yamada2018}. The demonstration of the global SU(4) symmetry employs SU(4) gauge transformations in the underlying Hubbard model. This result is remarkable given that  SU($N$) symmetries with larger values of $N$ are known to favor quantum  disordered states \cite{AffleckMarston1988,Arovas1988,ReadSachdev1989}. Furthermore,  a previous study showed compelling numerical evidence for a quantum spin-orbital liquid (QSOL) state in the SU(4) model on the honeycomb lattice \cite{Corboz2012}. However, in contrast with the Kitaev model, where the fractionalized excitations are Majorana fermions \cite{Kitaev2006}, the best candidate for the ground state of the SU(4) honeycomb model is a spin-orbital liquid described by a $\pi$-flux state of complex fermions at quarter filling  \cite{Corboz2012}.

In this paper, we provide an alternative derivation of the SU(4)-symmetric spin-orbital model for  $4d^1$ or $5d^1$ systems on the hyperhoneycomb lattice. The SU(4) symmetry of the  model is revealed by making use of a Klein transformation \cite{Kimchi2014,Kimchi2016} on the pseudospins. In addition, we derive the leading SU(4)-symmetry-breaking perturbations associated with Hund's coupling. Second, we investigate candidate spin-orbital liquid states using parton mean-field theories based on Majorana fermions or canonical (i.e. complex) fermions. We use these mean-field theories to construct trial wave functions, whose energies we evaluate after Gutzwiller projection using variational Monte Carlo (VMC) \cite{Gros1989}. Our results show that  the zero-flux state of complex fermions, which exhibits a spinon Fermi surface, has the lowest energy among the quantum spin-orbital liquids we consider. This contrasts with the result on the honeycomb lattice, where the $\pi$-flux state was energetically favored \cite{Corboz2012}.  Curiously, the $\pi$-flux state of complex fermions on the hyperhoneycomb displays three Dirac points. One of them has a spectrum with linear dispersion along two directions in  momentum  space and quadratic dispersion in the third direction. The other two points display the linear dispersion only along one direction. 

We also investigated possible instabilities of the zero-flux QSOL using a combination of VMC and linear flavor wave theory (LFWT). Within the SU(4)-symmetric model, a possible instability of the spin-orbital liquid is the formation of four-site SU(4) singlets \cite{Li1998}. The possible development of a state given by the direct product of four-site plaquettes, known as tetramerization, was systematically investigated by Ref. \cite{Lajko2013} on the honeycomb lattice. In this paper we demonstrate the stability of the zero-flux QSOL on the hyperhoneycomb lattice against tetramerization. We also studied the possibility of collinear long-range order formation due to perturbations induced by finite values of Hund's coupling. Within this set of ordered states, linear flavor wave theory (LFWT) \cite{Joshi1999,FranciscoKim2017} indicates that only a stripy ordered phase of $j=3/2$ moments is stable. However, further VMC computations showed that the QSOL is also stable against the formation of this order.

The paper is organized as follows. In Section \ref{derivation}, we derive the SU(4)-symmetric Hamiltonian from the multi-orbital Hubbard model in the limit of strong spin-orbit coupling. In Section \ref{meanfield}, we discuss the trial wave functions obtained by parton representations of the SU(4) generators. The energetics of these wave functions projected through VMC are presented in Section \ref{VMC}. Section \ref{sec:Hund} studies possible ordered phases induced by nonzero values of Hund's coupling. Finally, in Sec. \ref{sec:Conclusions} we offer some conclusions and suggestions for future developments. Technical details about the parton mean-field theories on the hyperhoneycomb lattice and LFWT can be found in the appendices.

\section{Effective spin-orbital model\label{derivation}}
We start from a multi-orbital Hubbard model for singly-occupied $4d$ or $5d$ orbitals in an octahedral crystal field. We  focus on the case where the edge-sharing octahedra form a hyperhoneycomb lattice [i.e. the $(10,3)b$ lattice \cite{OBrien2016}], but the derivation can be generalized to other tricoordinated lattices. We assume that the oxygen or halogen anions surrounding  the $d^1$ ion are in perfect octahedral arrangement.  The crystal field splits the $d$ levels into a lower-energy  $t_{2g}$ triplet ($|xy\rangle,|yz\rangle,|zx\rangle$) and a higher-energy    $e_g$ doublet. We can  label the $t_{2g}$ orbitals by the axis $\gamma=x,y,z$ perpendicular to the crystallographic plane containing them. For instance, $d_{jz\sigma}$ denotes the annihilation operator for an electron with spin $\sigma=\uparrow,\downarrow$ occupying the $xy$ orbital ($\gamma=z$) at site $j$. The multi-orbital Hubbard model is written as
\bea
\hspace{-0.4cm}H_{\text{Hub}} & =&-t\underset{ \gamma}{\sum}\underset{\langle ij\rangle_{\gamma}}{\sum} \underset{\sigma}{\sum} (d_{i\alpha\sigma}^{\dagger}d^{\phantom\dagger}_{j\beta\sigma}+d_{i\beta\sigma}^{\dagger}d^{\phantom\dagger}_{j\alpha\sigma}+\text{h.c.})\nonumber \\
 & & +\frac{1}{2}\underset{i}{\sum}\underset{\alpha\beta\alpha^{\prime}\beta^{\prime}}{\sum}\underset{\sigma\sigma^{\prime}}{\sum}U_{\alpha\beta;\alpha^{\prime}\beta^{\prime}}d_{i\alpha\sigma}^{\dagger}d_{i\beta\sigma^{\prime}}^{\dagger}d^{\phantom\dagger}_{i\beta^{\prime}\sigma^{\prime}}d^{\phantom\dagger}_{i\alpha^{\prime}\sigma}.\label{eq:Hubbard Hamiltonian}
\eea
In the first line of Eq.  (\ref{eq:Hubbard Hamiltonian}), $\langle ij\rangle_\gamma$ stands for a pair of nearest-neighbor sites connected by a bond in the  plane perpendicular to the $\gamma$ axis and  $\alpha,\beta$ are the other two spatial directions in the plane of the bond. %not equal to $\gamma$, such that $\alpha,\beta,\gamma$ is a cyclic permutation of $x,y,z$.  
This kinetic energy term  is bond- and orbital-dependent and takes into account only electron hopping via oxygen or halogen sites \cite{Jackeli2009}.  In the  interaction term, the parameters $U_{\alpha\beta;\alpha^{\prime}\beta^{\prime}}$ depend on matrix elements of the electrostatic potential between the $t_{2g}$ orbitals.  We keep only the dominant Coulomb terms, with the standard parametrization  $U_{\alpha\alpha;\alpha\alpha}\equiv U$ and $U_{\alpha\beta;\alpha\beta}\equiv U-2J_H$, where $J_H>0$ is   Hund's coupling constant \cite{khomskii2014transition}.

The magnetism of $4d$ and $5d$ compounds is strongly influenced by the atomic spin-orbit coupling. We then add to Hamiltonian  (\ref{eq:Hubbard Hamiltonian}) the   term\be
H_{\text{SOC}}=-\lambda\sum_j\mathbf l_j\cdot \mathbf S_j,
\ee
where $\lambda>0$ is the spin-orbit coupling constant, $\mathbf S_j$ is the electronic spin at site $j$, and $\mathbf l_j$ is the effective $l=1$ angular momentum of the $t_{2g}$ orbitals \cite{khomskii2014transition}. The spin-orbit coupling splits the  $t_{2g}$ levels into a $j=1/2$ doublet and a $j=3/2$ quadruplet, where $j$ is the quantum number associated with   $\mathbf J=\mathbf l+\mathbf S$.  The $j=3/2$ states have lower energy and are separated from the $j=1/2$ doublet by a gap $3\lambda/2$. In the limit $\lambda\gg t$, we can truncate the Hilbert space to the set of     $j=3/2$ states. It is convenient to represent the four states at each site in terms of two pseudospins $1/2$ as $|s^z,\tau^z\rangle$, with $s^z,\tau^z\in \left\{\frac12,-\frac12\right\}$, where $s^z$ is referred to as  the pseudospin eigenvalue and $\tau^z$ the pseudo-orbital eigenvalue \cite{Natori2016}.  We use the following convention for the local basis: \begin{align}
&\left|j^z=\frac32\right\rangle =\left|-\frac12,\frac12\right\rangle, &\left|j^z=\frac12\right\rangle =-\left|\frac12,-\frac12\right\rangle, \nonumber\\
&\left|j^z=-\frac12\right\rangle=\left|-\frac12,-\frac12\right\rangle, &\left|j^z=-\frac32\right\rangle =-\left|\frac12,\frac12\right\rangle.\end{align}
The convention is such that states with the same $\tau^z$   are conjugated by time reversal   and share  the same electronic density distribution \cite{Natori2017,comentario_base}. 

To  derive the effective spin-orbital model for the Mott insulating phase with $t\ll U$, we first consider $\lambda=0$ and   apply   perturbation theory to second order in $t/U$, imposing    the single-occupancy constraint $\sum_{\alpha,\sigma} d^\dagger_{j\alpha\sigma}d^{\phantom\dagger}_{j\alpha\sigma}=1$ \cite{Khaliullin2005}. Next, we take the limit of strong spin-orbit coupling   by projecting the Hamiltonian onto   $j=3/2$ states. The result is of the form $H_{\text{eff}}=\sum_{\langle ij\rangle_\gamma}H_{ij}^{(\gamma)}$ with
\begin{widetext}
%\bea
%H_{ij}^{(\gamma)}&=&\frac{J_{1}+J_{2}}{3}\left(2s_i^\gamma s_j^\gamma-\mathbf s_i\cdot \mathbf s_j+\frac14\right)\left(2\tau_i^y \tau_j^y-\boldsymbol \tau_i\cdot \boldsymbol \tau_j+\frac14\right)\nonumber\\
%&&+\frac{2}{9}\left(J_{2}-J_{3}\right)\left(2s_i^\gamma s_j^\gamma-\mathbf s_i\cdot \mathbf s_j-\frac14\right)\left(\boldsymbol \tau_i\cdot \boldsymbol \tau_j-\frac14\right)+\frac{J_{1}-J_{2}}{18} \left[Q_{i}^{\alpha\beta}%Q_{j}^{\alpha\beta}+2\left(\tau_{i}^{\beta\gamma}\tau_{j}^{\gamma\alpha}+\tau_{i}^{\gamma\alpha}\tau_{j}^{\beta\gamma}\right)\right]\nonumber\\
%&&-\frac{J_{1}-J_{2}}{30}\left[M_i^\gamma M_j^\gamma+2\mathbf M_i\cdot \mathbf M_j+\frac1{3}\left(  2M^\gamma_i    T^\gamma_j -\mathbf M_i\cdot \mathbf T_j +2T^\gamma_i    M^\gamma_j -\mathbf T_i\cdot \mathbf %M_j \right)\right].\label{bigH}
%\eea
\bea
 H_{ij}^{(\gamma)}& =&J_{a}\left[2\left(2s_{i}^{\gamma}s_{j}^{\gamma}-\mathbf{s}_{i}\cdot\mathbf{s}_{j}\right)+\frac{1}{2}\right]\left[2\left(2\tau_{i}^{y}\tau_{j}^{y}-\boldsymbol{\tau}_{i}\cdot\boldsymbol{\tau}_{j}\right)+\frac{1}{2}\right]\nonumber \\
 && +J_{b}\left[2\left(2s_{i}^{\gamma}s_{j}^{\gamma}-\mathbf{s}_{i}\cdot\mathbf{s}_{j}\right)-\frac{1}{2}\right]\left[2\boldsymbol{\tau}_{i}\cdot\boldsymbol{\tau}_{j}-\frac{1}{2}\right]+J_{c}\left[Q_{i}^{\alpha\beta}Q_{j}^{\alpha\beta}+2\left(\tau_{i}^{\beta\gamma}\tau_{j}^{\gamma\alpha}+\tau_{i}^{\gamma\alpha}\tau_{j}^{\beta\gamma}\right)\right]\nonumber \\
 && +\frac{J_{c}}{10}\left(-12M_{i}^{\gamma}M_{j}^{\gamma}-6\mathbf{M}_{i}\cdot\mathbf{M}_{j}+\mathbf{T}_{a,i}\cdot\mathbf{M}_{j}+\mathbf{M}_{i}\cdot\mathbf{T}_{a,j}-3T_{a,i}^{\gamma}M_{j}^{\gamma}-3M_{i}^{\gamma}T_{a,j}^{\gamma}\right)\nonumber \\
 && +\frac{\sqrt{15}J_{c}}{30}\left(T_{b,i}^{\alpha}M_{j}^{\alpha}+M_{i}^{\alpha}T_{b,j}^{\alpha}-T_{b,i}^{\beta}M_{j}^{\beta}-M_{i}^{\beta}T_{b,j}^{\beta}\right),\label{bigH}
\eea
\end{widetext}
where the coupling constants are
\bea
J_{a} & =&\frac{t^{2}}{3}\left(\frac{1}{U-3J_{H}}+\frac{1}{U-J_{H}}\right),\nonumber \\
J_{b} & =&\frac{2t^{2}}{9}\left(\frac{1}{U-J_{H}}-\frac{1}{U+2J_{H}}\right),\nonumber \\
J_{c} & =&\frac{2t^{2}}{9}\left(\frac{1}{U-3J_{H}}-\frac{1}{U-J_{H}}\right).
\eea
All the operators in Eq. (\ref{bigH}) are written in terms of components of $\mathbf s_j$ and $\boldsymbol \tau_j$, which act in the pseudospin and pseudo-orbital degree of freedom, respectively, and obey $[s_j^\alpha,s_{j'}^\beta]=i\delta_{jj'}\epsilon^{\alpha\beta\gamma}s^\gamma_j$, $[\tau_j^\alpha,\tau_{j'}^\beta]=i\delta_{jj'}\epsilon^{\alpha\beta\gamma}\tau^\gamma_j$, and $[s_j^\alpha,\tau_{j'}^\beta]=0$. The  15 operators $\{s^\alpha,\tau^\beta,s^\alpha\tau^\beta\}$ can be regarded as the generators of the SU(4) group. We define  $\tau^{\alpha\beta}$   as \cite{Natori2016}
\bea
\tau^{\alpha\beta}&=&u^\gamma_1\tau^z+u^\gamma_2\tau^x,\nonumber \\
\bar{\tau}^{\alpha\beta}&=&v^\gamma_1\tau^z+v^\gamma_2\tau^x,
\eea
where $\epsilon^{\alpha\beta\gamma}=1$ and we introduce the vectors $\mathbf u^\gamma\equiv (u^\gamma_1,u^\gamma_2)$ and $\mathbf v^\gamma\equiv (v^\gamma_1,v^\gamma_2)$ with  $\mathbf u^x=\left(-\frac12,\frac{\sqrt3}2\right)$,  $\mathbf u^y=\left(-\frac12,-\frac{\sqrt3}2\right)$, $\mathbf u^z=(1,0)$, $\mathbf v^x=\left(-\frac{\sqrt3}2,-\frac12\right)$, $\mathbf v^y=\left(\frac{\sqrt3}2,-\frac12\right)$, and $\mathbf v^z=(0,1)$. The other operators that appear in Eq. (\ref{bigH}) are given by \bea
M^\gamma&=&  -s^\gamma(1+4\tau^{\alpha\beta}), \label{eq:dipole}\\
T_{a}^{\gamma}&=&-3s^\gamma(1-\tau^{\alpha\beta}),\\
Q^{\alpha\beta}&=&-2\sqrt3s^\gamma \tau^y,\\
T_{b}^{\gamma}&=&-3\sqrt5 s^\gamma \bar{\tau}^{\alpha\beta}.
\eea
The vector $\mathbf M$ can be identified with the dipole  moment $\mathbf M=\mathbf J$ of the $j=3/2$ multiplet \cite{Chen2010,Natori2017}. Similarly, $T^\gamma_{a}$ is an octupole forming a $\Gamma_{4}$ irreducible representation of the octahedral group. The $Q^{\alpha\beta}$ and $T^\gamma_{b}$ correspond, respectively, to quadrupole and octupole moments forming a $\Gamma_{5}$ irreducible representation.  

In general, the effective Hamiltonian   (\ref{bigH}) is invariant under space group  transformations (Fddd for the hyperhoneycomb lattice), but lacks any continuous symmetry, as expected for spin-orbit-coupled systems. The general result is greatly simplified if we take the limit of vanishing  Hund's coupling. Seting $J_H=0$, we obtain $H_{ij}^{(\gamma)}\to \bar H_{ij}^{(\gamma)}$, where 
\bea
\bar H_{ij}^{(\gamma)}&=&J\underset{\langle ij\rangle_{\gamma}}{\sum}\left[2\left(2s_{i}^{\gamma}s_{j}^{\gamma}-\mathbf{s}_{i}\cdot\mathbf{s}_{j}\right)+\frac{1}{2}\right]\nonumber \\
&&\quad \,\,\,\,\times\left[2\left(2\tau_{i}^{y}\tau_{j}^{y}-\boldsymbol{\tau}_{i}\cdot\boldsymbol{\tau}_{j}\right)+\frac{1}{2}\right],\label{HJHeq0}
\eea
with $J=2t^2/(3U)$. 

The coupling between pseudospins $\mathbf s$ in Eq. (\ref{HJHeq0}) is reminiscent of a special point of the Kitaev-Heisenberg model where the ground state is  known  exactly \cite{Chaloupka2010,Kimchi2014,Lee2014}. This observation suggests performing a four-sublattice  rotation on the pseudospins. Such rotations have been called Klein dualities in Ref. \cite{Kimchi2014} because the set of transformations is isomorphic to the Klein four-group $\mathbb Z_2\times \mathbb Z_2$. %In our case, the Klein transformation must act differently on  $\mathbf s$ and $\boldsymbol \tau$. 
Conveniently, the hyperhoneycomb lattice  can be viewed as a face-centered orthorrombic lattice with a four-point basis \cite{Lee2014,Takayama2015}.  Let us denote the sublattices by $A_r$, with $r=1,\dots,4$.  We define the  Klein transformation   \bea
\tilde{\textbf{s}}_{i}&=&\begin{cases}
\textbf{s}_{i}, & i\in A_1,\\
(-s_{i}^{x},-s_{i}^{y},s_{i}^{z}), & i\in A_2,\\
(s_{i}^{x},-s_{i}^{y},-s_{i}^{z}), & i\in A_3,\\
(-s_{i}^{x},s_{i}^{y},-s_{i}^{z}), & i\in A_4.
\end{cases}\label{eq:Klein transformation}
\eea
This transformation is such that, for any bond $\langle ij\rangle_\gamma$,\be
2s_i^\gamma s_j^\gamma-\mathbf s_i\cdot\mathbf s_j=\tilde{\mathbf s}_i\cdot\tilde{\mathbf s}_j.
\ee
On the other hand, the pseudo-orbital coupling in Eq. (\ref{HJHeq0})  is bond independent. We define \bea
\tilde{\boldsymbol{\tau}}_{i}&=&\begin{cases}
\boldsymbol{\tau}_{i}, &  i \in A_r \text{ with $r$ even},\\
(-\tau_{i}^{x},\tau_{i}^{y},-\tau_{i}^{z}), & i \in A_r \text{ with $r$ odd}.
\end{cases}\label{eq:pseudo-orbital transformation}
\eea
This is such that, for any bipartite lattice, \be
2\tau_i^y \tau_j^y-\boldsymbol \tau_i\cdot\boldsymbol \tau_j=\tilde{\boldsymbol \tau}_i\cdot\tilde{\boldsymbol \tau}_j. \label{eq:two-sublattice transformation}
\ee
Note that $\tilde {\mathbf s}$ and $\tilde {\boldsymbol \tau}$ obey the same algebra as the original $  {\mathbf s}$ and $  {\boldsymbol \tau}$. Applying the transformations in Eqs.    (\ref{eq:Klein transformation}) and (\ref{eq:pseudo-orbital transformation}), we find that the effective Hamiltonian for ${J_H=0}$ becomes 
\be
\bar H_{\text{eff}}=J\underset{\langle ij\rangle}{\sum}\left(2\tilde{\textbf{s}}_{i}\cdot\tilde{\textbf{s}}_{j}+\frac{1}{2}\right)\left(2\tilde{\boldsymbol{\tau}}_{i}\cdot\tilde{\boldsymbol{\tau}}_{j}+\frac{1}{2}\right).\label{eq:SU(4) final}
\ee
This is the familiar form of SU(4)-symmetric spin-orbital models as studied, for instance, in  Refs. \cite{Li1998,Pati1998,Itoi2000,Wang2009,Lajko2013,Kugel2015}. We stress, however, that these previous studies were motivated by systems with doubly degenerate orbitals. Here we started with triply degenerate $t_{2g}$ orbitals and the strong spin-orbit coupling plays an essential role in the emergence of the SU(4) symmetry in the $j=3/2$ subspace. Moreover, the conserved quantities are not associated with the total spin and orbital angular momentum, but rather with  the rotated pseudospin and pseudo-orbital operators   $\sum_i\tilde s_i^\alpha,\sum_i\tilde \tau_i^\beta,\sum_i\tilde s_i^\alpha  \tilde\tau_i^\beta$. 

One advantage of our derivation based on Klein transformations is that it provides a simple criterion  to verify whether the spin-orbital model on a given lattice presents or not an emergent SU(4) symmetry.  In fact, it has been shown  \cite{Kimchi2014} that if $N_{x}(p), N_{y}(p), N_{z}(p)$ are, respectively, the number of $x, y, z$ bonds in a given plaquette $p$ of the lattice,  the Klein transformation in Eq. (\ref{eq:Klein transformation}) can be defined if and only if 
$N_{x}(p), N_{y}(p), N_{z}(p)$ are  either all even or all odd for all   plaquettes.  In addition, the pseudo-orbital rotation in Eq. (\ref{eq:pseudo-orbital transformation}) requires that the lattice be bipartite.  All the tricoordinated lattices studied in Ref. \cite{Yamada2018} satisfy these constraints. By contrast, the triangular lattice can be built from edge-sharing octahedra \cite{Jackeli2009}, but in this case the model (\ref{HJHeq0}) cannot be cast in the SU(4)-symmetric form of Eq. (\ref{eq:SU(4) final}) because the triangular lattice is not bipartite. 

\begin{figure}
\begin{centering}
\includegraphics[width=0.9\columnwidth]{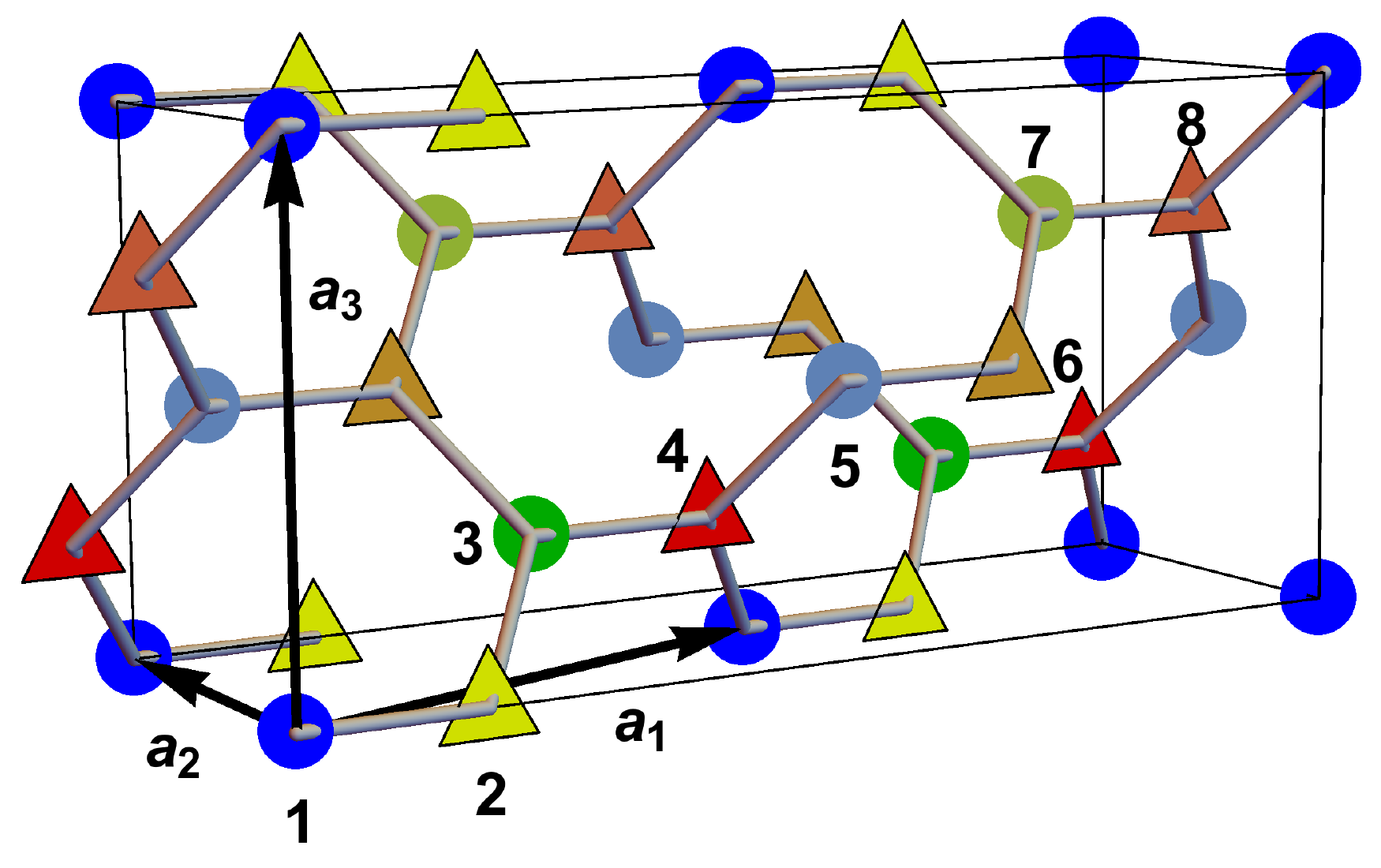}
\par\end{centering}
\caption{\label{fig:hyperhoneycomb lattice} (Color online) Hyperhoneycomb
lattice   as a base-centered orthorhombic lattice with an  eight-point basis. The disks and
triangles indicate that the lattice is bipartite, while the
different colors represent the different sublattices $r=1,\dots,8$. The primitive lattice vectors $\mathbf a_1$, $\mathbf a_2$, $\mathbf a_3$  (see Appendix \ref{sec:Hyperhoneycomb}) are also shown.}
\end{figure}

Although the hyperhoneycomb lattice admits a four-sublattice decomposition, for the purposes of Section \ref{meanfield} it will be convenient to double the unit cell and consider a base-centered orthorhombic lattice with an eight-point basis. The eight sublattices are illustrated in Fig. \ref{fig:hyperhoneycomb lattice}. In this case, we simply extend  Eq. (\ref{eq:Klein transformation}) such that the transformation on the pseudospins on sublattices $A_5$, $A_6$, $A_7$ and $A_8$ correspond to the transformations on sublattices $A_2$, $A_1$, $A_4$ and $A_3$, respectively \cite{Lee2014}. On the other hand, the pseudo-orbital transformations are still defined by the parity of the sublattices in accordance to Eq. (\ref{eq:two-sublattice transformation}).  
 
Going back to Eq. (\ref{bigH}), we can rewrite  the complete Hamiltonian for $J_H\neq0$ in terms of the rotated operators. Let us first define the three auxiliary Hamiltonians
%\begin{widetext}\bea
%\bar H_{ij}^{(\gamma)}&=&(J_a-J_b)\left(2\tilde{\textbf{s}}_{i}\cdot\tilde{\textbf{s}}_{j}+\frac{1}{2}\right)\left(2\tilde{\boldsymbol{\tau}}_{i}\cdot\tilde{\boldsymbol{\tau}}_{j}+\frac{1}{2}\right)+8J_b  \tilde{\textbf{s}}_{i}\cdot\tilde{\textbf{s}}_{j}\tilde{\tau}^y_i\tilde{\tau}^y_j+J_b\left(2\tilde{\boldsymbol\tau}_i\cdot \tilde{\boldsymbol\tau}_j-\frac12\right)-2J_b\tilde{\tau}^y_i\tilde{\tau}^y_j\nonumber\\
%&&+2J_c\left(6\tilde{s}^\gamma_i\tilde{s}^\gamma_j\tilde{\tau}^y_i\tilde{\tau}^y_j-\tilde{\tau}_i^{\beta\gamma}\tilde{\tau}_j^{\gamma\alpha}-\tilde{\tau}_i^{\gamma\alpha}\tilde{\tau}_j^{\beta\gamma}\right)-3J_c \tilde{s}%^\gamma_i\tilde{s}^\gamma_j\nonumber\\
%&&-3J_c\left[3\tilde{s}^\gamma_i\tilde{s}^\gamma_j\left(\tilde{\tau}^{\alpha\beta}_i-\tilde{\tau}^{\alpha\beta}_j\right)-\tilde{s}^\alpha_i\tilde{s}^\alpha_j\left(\tilde{\tau}^{\beta\gamma}_i-\tilde{\tau}^{\beta\gamma}_j\right)-\tilde{s}^\beta_i\tilde{s}^\beta_j\left(\tilde{\tau}^{\gamma\alpha}_i-\tilde{\tau}^{\gamma\alpha}_j\right)\right]\nonumber\\
%&&+24J_c\left(\tilde{s}^\gamma_i\tilde{\tau}^{\alpha\beta}_i\tilde{s}^\gamma_j\tilde{\tau}^{\alpha\beta}_j-\tilde{s}^{\alpha}_i\tilde{\tau}^{\beta\gamma}_i\tilde{s}^{\alpha}_j\tilde{\tau}^{\beta\gamma}_j-\tilde{s}^{\beta}_i\tilde{\tau}^{\gamma\alpha}_i\tilde{s}^{\beta}_j\tilde{\tau}^{\gamma\alpha}_j\right).\label{bigH2}
%\eea
%\end{widetext}
\bea
\mathcal{H}_{\text{SU(4)},ij} & =&\left(2\tilde{\mathbf{s}}_{i}\cdot\tilde{\mathbf{s}}_{j}+\frac{1}{2}\right)\left(2\tilde{\boldsymbol{\tau}}_{i}\cdot\tilde{\boldsymbol{\tau}}_{j}+\frac{1}{2}\right), \\ \label{HSU4temp}
\mathcal{H}_{b,ij} & =&8\tilde{\mathbf{s}}_{i}\cdot\tilde{\mathbf{s}}_{j}\tilde{\tau}_{i}^{y}\tilde{\tau}_{j}^{y}+2(\tilde{\tau}_{i}^{x}\tilde{\tau}_{j}^{x}+\tilde{\tau}_{i}^{z}\tilde{\tau}_{j}^{z})+\frac{1}{2}, \\ \label{Hbtemp}
\mathcal{H}_{c,ij}^{(\gamma)} & =& 3\tilde{s}_{i}^{\gamma}\tilde{s}_{j}^{\gamma}\left[4\tilde{\tau}_{i}^{y}\tilde{\tau}_{j}^{y}+8\tilde{\tau}_{i}^{\alpha\beta}\tilde{\tau}_{j}^{\alpha\beta}-3(\tilde{\tau}_{i}^{\alpha\beta}-\tilde{\tau}_{j}^{\alpha\beta})\right] \nonumber \\
&& -3\tilde{s}_{i}^{\gamma}\tilde{s}_{j}^{\gamma}\nonumber \\ &&-8\left(\tilde{s}_{i}^{\alpha}\tilde{s}_{j}^{\alpha}\tilde{\tau}_{i}^{\beta\gamma}\tilde{\tau}_{j}^{\beta\gamma}+\tilde{s}_{i}^{\beta}\tilde{s}_{j}^{\beta}\tilde{\tau}_{i}^{\gamma\alpha}\tilde{\tau}_{j}^{\gamma\alpha}\right)\nonumber\\
&&+4\left(\tilde{s}_{i}^{\alpha}\tilde{s}_{j}^{\alpha}+\tilde{s}_{i}^{\beta}\tilde{s}_{j}^{\beta}-\frac{1}{2}\right)\left(\tilde{\tau}_{i}^{\beta\gamma}\tilde{\tau}_{j}^{\gamma\alpha}+\tilde{\tau}_{i}^{\gamma\alpha}\tilde{\tau}_{j}^{\beta\gamma}\right)\nonumber\\
&&+\sqrt{3}\left(\tilde{s}_{i}^{\alpha}\tilde{s}_{j}^{\alpha}-\tilde{s}_{i}^{\beta}\tilde{s}_{j}^{\beta}\right)\left(\tilde{\bar{\tau}}_{i}^{\alpha\beta}-\tilde{\bar{\tau}}_{j}^{\alpha\beta}\right)\label{Hctemp}.
\eea
The complete Hamiltonian then reads 
\be
\bar{H}_{ij}^{(\gamma)}=(J_{a}-J_{b})\mathcal{H}_{\text{SU(4)},ij}+J_{b}\,\mathcal{H}_{b,ij}+J_{c}\,\mathcal{H}_{c,ij}^{(\gamma)}.\label{bigH2}
\ee
It is then clear that the SU(4) symmetry is lost once $J_b$ and $J_c$ are nonzero. In Section \ref{meanfield}, we shall focus on the SU(4)-symmetric model, but we will return to the question about the effects of finite Hund's coupling in Section \ref{sec:Hund}. 
 
\section{Candidate spin liquid states at the SU(4)-symmetric point\label{meanfield}}

Inspired by the numerical evidence for a quantum spin-orbital liquid in the SU(4) model on the honeycomb lattice \cite{Corboz2012}, in this section we investigate fermionic parton mean-field theories for the model on the hyperhoneycomb lattice. While we cannot rule out a symmetry-breaking ground state, the study of quantum spin-orbital liquids  will be justified  a posteriori in Subsection \ref{VMC} by showing that the corresponding  variational states for the hyperhoneycomb model have energies comparable to those in the honeycomb model and that they are stable against perturbations such as tetramerization. 

\subsection{Parton mean-field theory}
We start with the representation that employs canonical complex fermions \cite{ReadSachdev1989,Li1998,Joshi1999}. First, we rewrite the four states in the local basis $|\tilde s^z,\tilde \tau^z\rangle$ as
\begin{align}
|1\rangle&=\left|\frac{1}{2},\frac{1}{2}\right\rangle , \quad &|2\rangle&=\left|-\frac{1}{2},\frac{1}{2}\right\rangle ,\nonumber \\
|3\rangle&=\left|\frac{1}{2},-\frac{1}{2}\right\rangle , \quad &|4\rangle&=\left|-\frac{1}{2},-\frac{1}{2}\right\rangle .\label{colors}
\end{align}
With this notation, we can define the generators of the   SU(4) group \be
S_m^n=|m\rangle \langle n|, \qquad m,n=1\dots, 4, 
\label{generators}\ee
which obey the algebra\be
[S_m^n,S_{m'}^{n'}]= \delta_{n,m'}S_m^{n'}-\delta_{m,n'}S_{m'}^n.
\ee
On the lattice, we define local generators $S_m^n(i)$ at each site $i$, which obey \be
[S_m^n(i),S_{m'}^{n'}(j)]=\delta_{ij}\delta_{n m'}S_m^{n'}(i)-\delta_{ij}\delta_{m n'}S_{m'}^n(i).\label{localalgebra}\ee  
In terms of the local SU(4) generators, the Hamiltonian in Eq. (\ref{eq:SU(4) final}) can be written as\be
\bar H_{\text{eff}}=J\sum_{\langle ij\rangle}\sum_{m,n=1}^4S_m^n(i)S_n^m(j).\label{Hferm}
\ee

We now introduce fermion creation operators $f^\dagger_m$, with four ``colors'' $m=1,\dots,4$ \cite{Corboz2012}, by\be
|m\rangle = f^\dagger_m| \emptyset\rangle,
\ee
where $| \emptyset\rangle$ is the vacuum of the Fock space. The SU(4) generators for each site $j$ are represented by  \be
S_m^n(j)=f^\dagger_{jm}f^{\phantom\dagger}_{jn}.\label{Su4gen}
\ee
The physical states obey the single-occupancy constraint \be
\sum_m f^\dagger_{jm}f^{\phantom\dagger}_{jm}=1\quad\forall j.\label{U1constraint}
\ee 
It follows from   canonical anticommutation relations, $\{f^{\phantom\dagger}_{im},f^{ \dagger}_{jn}\}=\delta_{ij}\delta_{mn}$, that the operators in Eq. (\ref{Su4gen}) obey  the algebra in Eq. (\ref{localalgebra}). 
This provides a fundamental representation of SU(4) in terms of a four-component fermionic spinor $(f_{j1},f_{j2},f_{j3},f_{j4})^T$.

While the Hamiltonian in Eq. (\ref{Hferm}) is quartic in the fermion operators, a quadratic Hamiltonian can be obtained using  a  decoupling with symmetry-preserving  parameters $\langle f^{ \dagger}_{im} f^{\phantom\dagger}_{jm} \rangle$ \cite{Zhou2017}. We then consider the mean-field  Hamiltonian\be
H_f=-\sum_{\langle ij \rangle}\sum_{m=1}^4(\chi_{ij} f^{ \dagger}_{im} f^{\phantom\dagger}_{jm}+\text{h.c.}), \label{HMF}
\ee
where $\chi_{ij}$ are the mean-field parameters that specify the spin liquid ansatz.  This kind of mean-field decoupling becomes exact, for instance, in the case of the self-adjoint representation (with $N/2$ fermions per site for $N$ even) of SU($N$) in the limit $N\to\infty$ \cite{AffleckMarston1988,ReadSachdev1989}.  In this limit,    a saddle-point approximation in the fermionic action is justified and fluctuations of the emergent gauge field can be neglected, rendering the fermions noninteracting.
The ground state in this limit does not break the SU($N$) symmetry (as oppposed to N\'eel-type states) and  can be either a valence bond solid or a quantum spin liquid. More generally, the mean-field decoupling leading to Eq. (\ref{HMF}) has been   used to generate variational wave functions for SU($N$) models with finite   $N$, for instance for the $N=4$ model on the honeycomb lattice \cite{Corboz2012}. The validity of such wave functions as approximations for the true ground state has to be tested numerically by computing their corresponding    energies \cite{Paramekanti2007}.

\begin{figure}
\centering{}\subfloat[\label{fig:zero flux Majorana}]{\includegraphics[width=0.35\columnwidth]{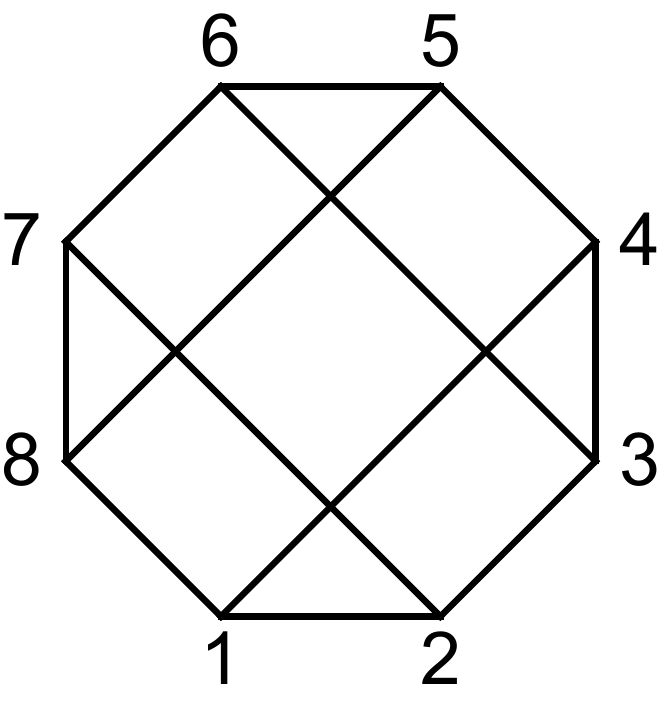}

}\qquad\quad \subfloat[\label{fig:pi flux Majorana}]{\includegraphics[width=0.35\columnwidth]{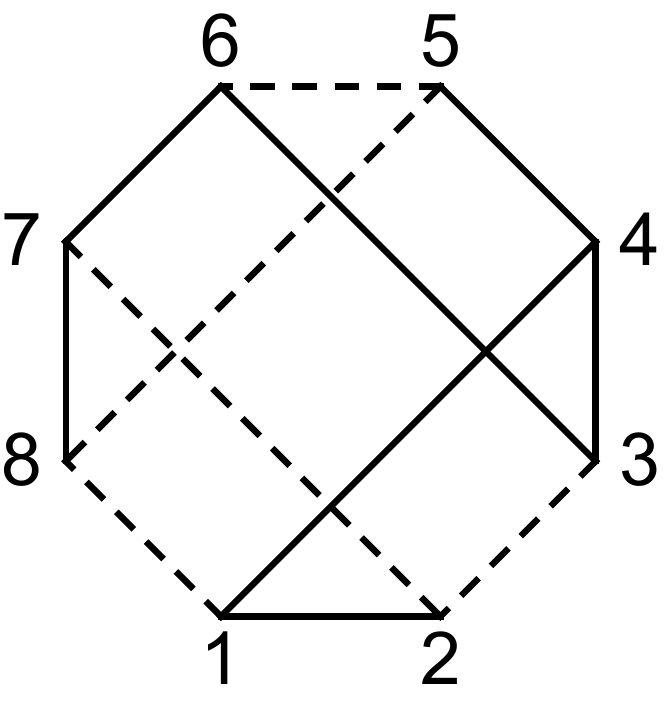}

}
\caption{\label{fig:ansatze}Representation of the (a) zero-flux and 
(b) $\pi$-flux states. Each vertex corresponds
to a basis point of the hyperhoneycomb lattice as labeled in Fig.
\ref{fig:hyperhoneycomb lattice}. Solid (dashed) lines represent bonds with $\chi_{ij}=+1$ ($\chi_{ij}=-1$).}
\end{figure}

The hermiticity of $H_f$ imposes $\chi_{ij}=\chi_{ji}^*$. Following Ref. \cite{Corboz2012}, we consider Ans\"atze that preserve  SU(4) as well as time reversal and crystalline symmetries and restrict ourselves to $\chi_{ij}\in \mathbb R$.   Fixing $\chi_{ij}=\pm 1$, we can label physical states  by the gauge-invariant fluxes \be
e^{i\Phi(P)}=\prod_{\langle ij \rangle\in P} \chi_{ij},
\ee
where $P$ is a 10-site elementary loop on the hyperhoneycomb lattice (see Fig. \ref{fig:hyperhoneycomb lattice} and Appendix \ref{sec:Hyperhoneycomb}). There are two states with uniform flux through all loops (see Fig. \ref{fig:ansatze}). The zero-flux state [$\Phi(P)=0\;\forall P$] can be described by assigning $\chi_{ij}=+1$ to all bonds. The $\pi$-flux state  [$\Phi(P)=\pi\;\forall P$]  is obtained by setting $\chi_{ij}=+1$ on the bonds represented by solid lines in Fig. \ref{fig:pi flux Majorana} and $\chi_{ij}=-1$ on those represented by dashed lines. While the zero-flux state could be represented using four sublattices, the $\pi$-flux state requires the eight-sublattice representation   of the hyperhoneycomb lattice.

At the mean-field level, the single-occupancy constraint is imposed on average, $\sum_m\langle \Psi_{\text{MF}}| f^\dagger_{im}f^{\phantom\dagger}_{im} |\Psi_\text{MF}\rangle=1$, corresponding to a quarter-filled Fermi sea. We can determine the mean-field ground state $|\Psi_{\text{MF}}\rangle$ by diagonalizing the quadratic Hamiltonian (\ref{HMF}) for both zero-flux and $\pi$-flux states.  We   obtain\be
H_f=\sum_{\mathbf k}\sum_{\lambda=1}^8E_{\lambda}(\mathbf k)f^\dagger_{\mathbf k\lambda}f^{\phantom\dagger}_{\mathbf k\lambda},
\ee 
where $\lambda$ is the band index and $f_{\mathbf k\lambda}$ annihilates a fermion with momentum $\mathbf k$ in band $\lambda$. For the zero-flux state, we have  analytical expressions for the dispersion relations. They can be written as $E_1=\mc E_1^{++}$, $E_2=\mc E_1^{+-}$, $E_3=\mc E_1^{-+}$, $E_4=\mc E_1^{--}$, $E_5=\mc E_2^{++}$, $E_6=\mc E_2^{+-}$, $E_7=\mc E_2^{-+}$, $E_8=\mc E_2^{--}$, where  \bea
\mc E_{n}^{pp'}(\mathbf k)=& p\sqrt{g_n(\mathbf k)+p'\sqrt{[g_n(\mathbf k)]^2-|h_n(\mathbf k)|^2}},\label{dispersion}
\eea
with  $n=1,2$ and $p,p'=\pm$, and we define the  functions
\bea
g_{1}(\mathbf{k}) & =&3+2\cos(2k_{z})\cos(k_{x}-k_{y}),\nonumber \\
h_{1}(\mathbf{k}) & =&2\cos(2k_{z})e^{i(k_{x}+k_{y})}-e^{-i2(k_{x}+k_{y})}\nonumber \\
 && +e^{i2k_{x}}+e^{i2k_{y}},\nonumber \\
g_{2}(\mathbf{k}) & =&3-2\cos(2k_{z})\cos(k_{x}-k_{y}),\nonumber \\
h_{2}(\mathbf{k}) & =&2\cos(2k_{z})e^{i(k_{x}+k_{y})}+e^{-i2(k_{x}+k_{y})}\nonumber \\
 && -e^{i2k_{x}}-e^{i2k_{y}}.\label{eq:fi and gi}
\eea
For the $\pi$-flux state, we were only able to find the dispersion relations numerically. More details are provided in Appendix \ref{sec:Details MFT}. Figure \ref{fig:Mean-field-dispersion} shows the dispersions for both zero-flux and $\pi$-flux states. We note that the dispersions are particle-hole symmetric, as expected  since the lattice is bipartite.

\begin{figure}
\begin{centering}
\subfloat[\label{fig:zero flux dispersion}]{\begin{centering}
\includegraphics[width=.9\columnwidth]{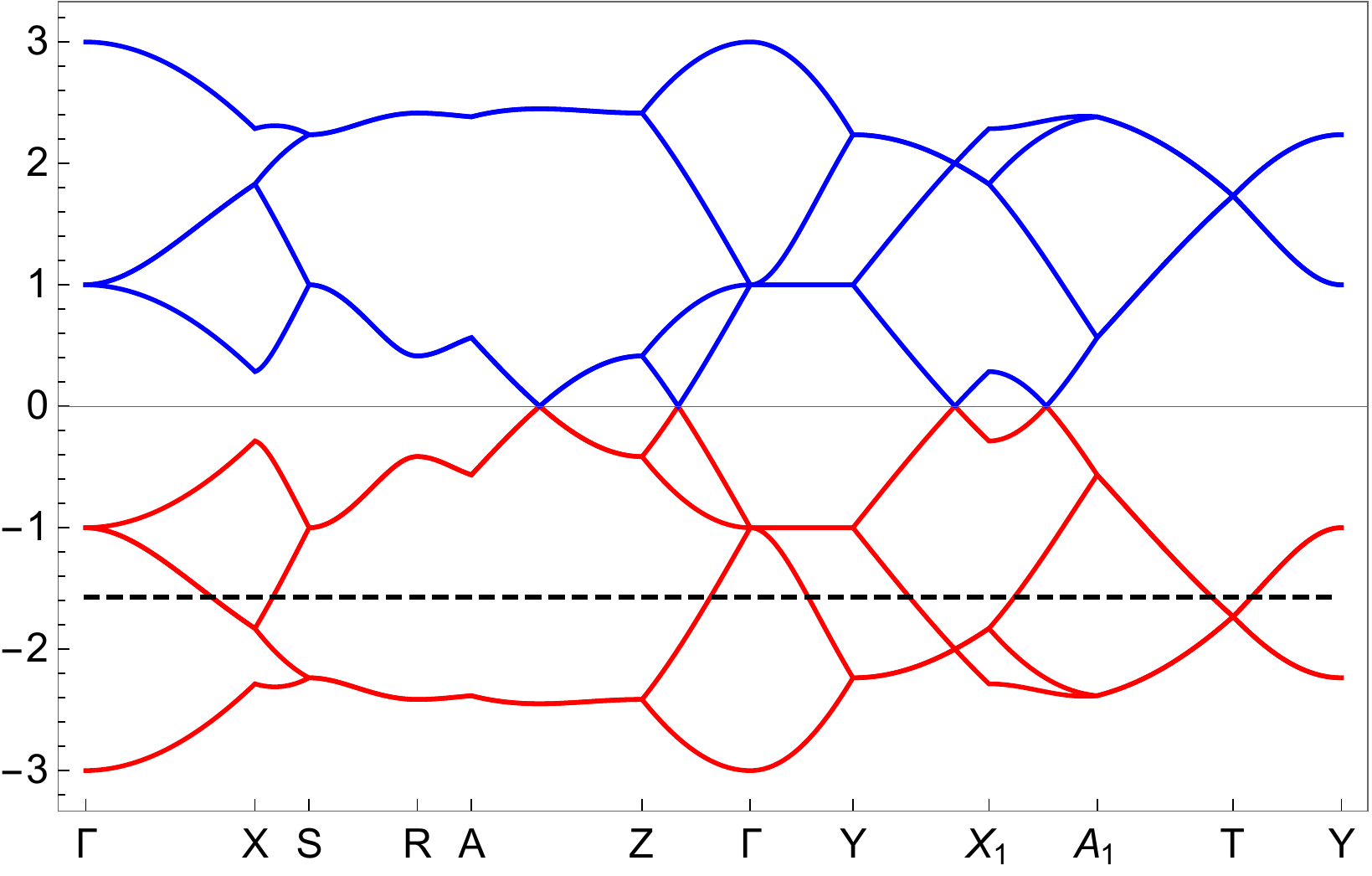}
\par\end{centering}

}
\par\end{centering}

\begin{centering}
\subfloat[\label{fig:pi-flux dispersion}]{\begin{centering}
\includegraphics[width=.9\columnwidth]{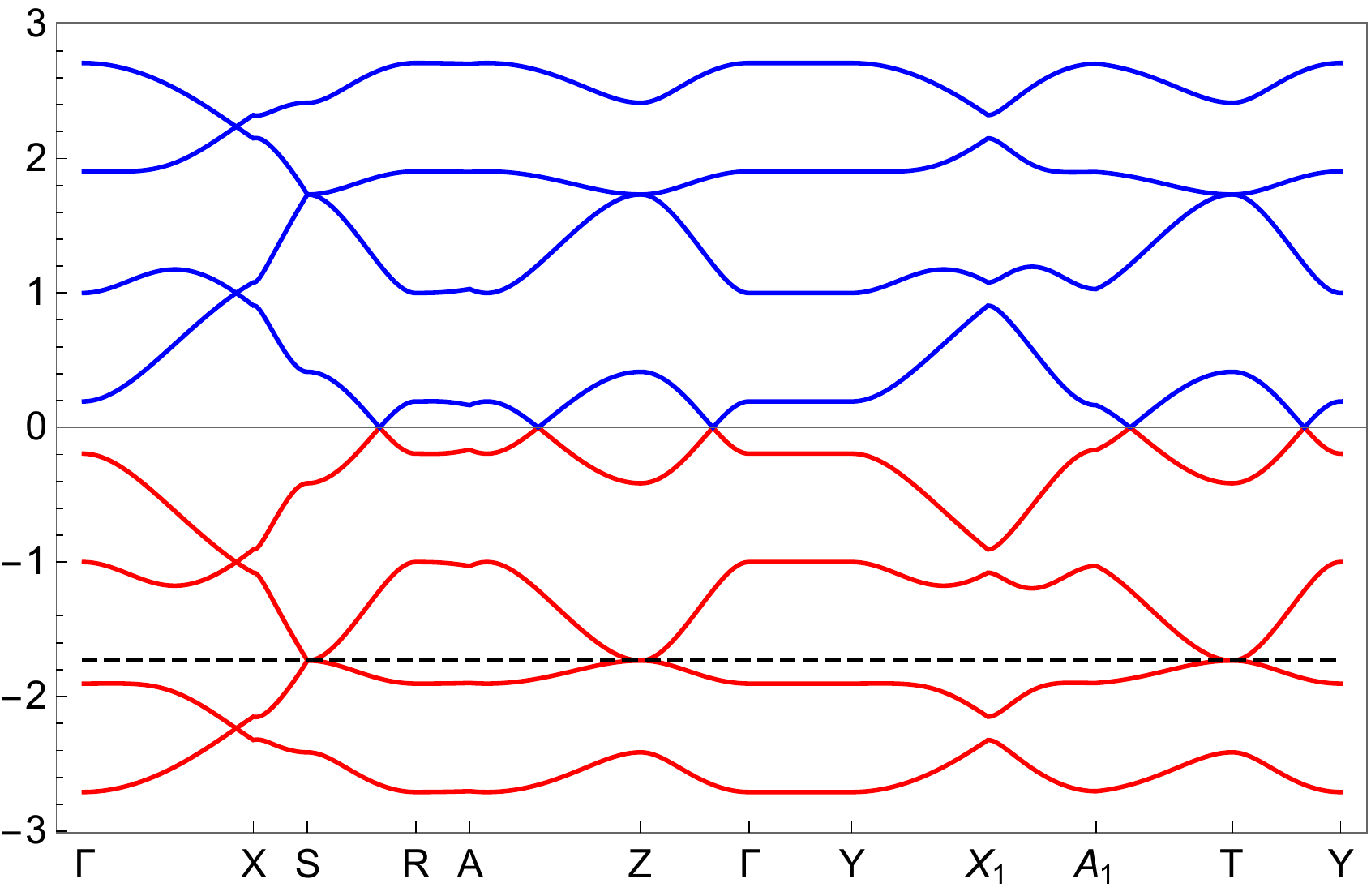}
\par\end{centering}

}
\par\end{centering}

\caption{\label{fig:Mean-field-dispersion}(Color online) Mean-field  dispersion of fermions in the  (a) zero-flux and (b) $\pi$-flux states. The dashed line marks the   Fermi level at quarter filling. The high symmetry points of the Brillouin zone      are  specified in Fig. \ref{fig:Fermi surfaces}. The energy scale in this plot is set by $|\chi_{ij}|=1$.  }
\end{figure}

\begin{figure}
\begin{centering}
\subfloat[\label{fig:Dirac zero-flux Fermi Surface}]{\begin{centering}
\includegraphics[width=0.55\columnwidth]{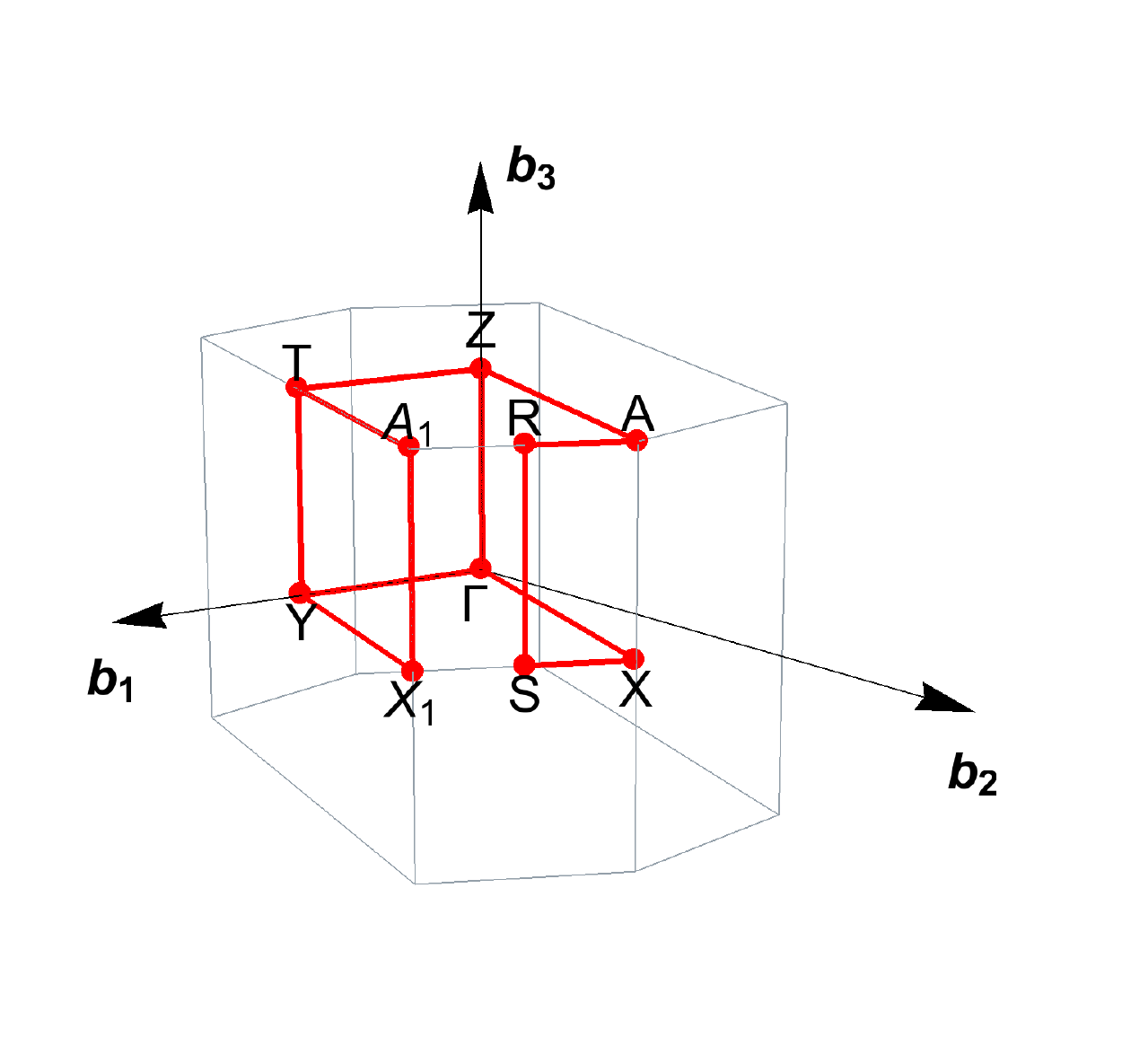}
\par\end{centering}

}\subfloat[\label{fig:Dirac pi flux semi-Dirac points}]{\includegraphics[width=0.4\columnwidth]{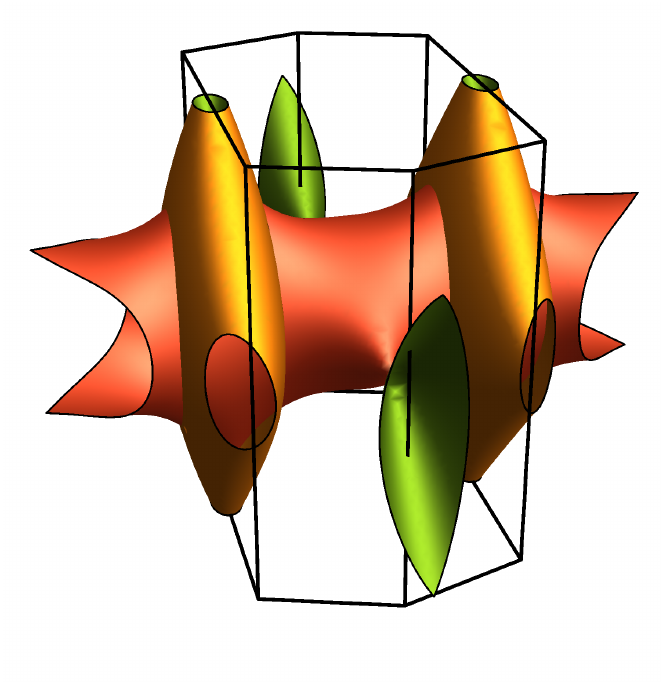}

}
\par\end{centering}

\caption{\label{fig:Fermi surfaces}(Color online) (a) Brillouin zone of the base-centered orthorhombic lattice. (b) Fermi surface  of the zero-flux state inside the Brillouin zone. Different colors represent different bands in Eq. (\ref{dispersion}). }
\end{figure}

The quarter-filling condition $\langle f^\dagger_{im}f^{\phantom\dagger}_{im} \rangle=1/4$ determines  the position of the Fermi level. The zero-flux state displays a Fermi surface   illustrated in Fig.  \ref{fig:Fermi surfaces}. The two pieces of Fermi surface depicted in yellow are connected by the vector $\mathbf Q_0=\left(\frac\pi3,\frac\pi3,\frac\pi3\right)$. This is a reciprocal lattice vector of the face-centered orthorrombic lattice, i.e., the Bravais lattice of the hyperhoneycomb lattice before doubling the unit cell. Thus, the Fermi surface is not nested and this quantum spin-orbital liquid is at least locally stable against (spin) density waves driven by interactions beyond the mean-field level. On the other hand, for the $\pi$-flux state the Fermi level crosses Dirac points at the high-symmetry points S, Z and T.  Close inspection  reveals that the dispersion in the vicinity of these Dirac points is anisotropic. The spectrum in the neighborhood of S is linear along two directions in $\mathbf k$ space but quadratic in the third direction. The opposite is verified for the dispersion around the Z and T points, which is quadratic along two directions and linear in the third. Similar behavior has been discussed for Dirac semimetals in two and three dimensions  \cite{Dietl2008,Banerjee2009,Yang2014}.

The mean-field ground state $|\Psi_\text{MF}\rangle$ for zero-flux and $\pi$-flux states correspond to occupying all the single-fermion states with   energy $E_\lambda(\mathbf k)$ below the Fermi level.  We obtain a  variational wave function in the physical Hilbert space by imposing the local single-occupancy constraint (\ref{U1constraint}) via the Gutzwiller projection \be
|\Psi_{\text{phys}}\rangle=\mc P_f|\Psi_\text{MF}\rangle,
\ee
where $\mc P_f=\prod_i\left[ \frac16n_i (2-n_i)(3-n_i)(4-n_i)\right]$ with $n_i=\sum_mf^\dagger_{im}f^{\phantom\dagger}_{im}$. In practice, the Gutzwiller projection is implemented numerically on finite lattices  using VMC, as we shall discuss in Section \ref{VMC}.

Let us now discuss the parton mean-field theory generated by a Majorana fermion representation of pseudospin and pseudo-orbital operators \cite{Wang2009, Natori2016}. Using SU(4)$\cong$ SO(6), we can represent the SU(4) generators using 6 Majorana fermions $\{\eta^\gamma,\theta^\gamma\}$, with $\gamma=x,y,z$, in the form\bea
\tilde s^\gamma&=&-\frac{i}4\epsilon^{\alpha\beta\gamma}\eta^\alpha\eta^\beta,\\
\tilde \tau^\gamma&=&-\frac{i}4\epsilon^{\alpha\beta\gamma}\theta^\alpha\theta^\beta.
\eea
The Majorana fermion operators  obey the anticommutation  relations $\{\eta_j^\alpha,\eta_{j'}^\beta\}=2\delta_{jj'}\delta^{\alpha\beta}=\{\theta_j^\alpha,\theta_{j'}^\beta\}$ and $\{\eta_j^\alpha,\theta_{j'}^\beta\}=0$. To deal with the $\mathbb Z_2$ gauge redundancy  of this representation, we must impose the local constraint\be
i\eta^x_j\eta^y_j\eta^z_j\theta^x_j\theta^y_j\theta^z_j=1\quad \forall j. \label{Z2constraint}
\ee
It follows from Eq. (\ref{Z2constraint}) that $\tilde s^\alpha\tilde \tau^\beta=-\frac{i}4\eta^\alpha\theta^\beta$. Thus, all the SU(4) generators are quadratic in Majorana fermions. 
 
It is convenient to construct three complex fermions from the Majorana fermions  as $c_{j\gamma}=(\eta_j^\gamma-i\theta_j^\gamma)/2$.  In terms of the three-component vector $\mathbf c_j=(c_{jx},c_{jy},c_{jz})$, the model in Eq. (\ref{eq:SU(4) final}) reads \cite{Wang2009}\be
\bar H_{\text{eff}}=J\sum_{\langle jl\rangle}\left[1-\frac{1}{2}\left(i  \mathbf{c}_{j}^{\dagger}\cdot   \mathbf{c}^{\phantom\dagger}_{l }- i\mathbf{c}_{l }^{\dagger}  \cdot\mathbf{c}^{\phantom\dagger}_{j }\right)^{2}\right].
\ee
The constraint in Eq. (\ref{Z2constraint}) can be written as \be
\prod_{\gamma=x,y,z}\left(1-2{c}_{j\gamma}^{\dagger}  {c}^{\phantom\dagger}_{j\gamma}\right)=1\quad\forall j.
\ee
In other words, the physical states are those with an even number of $c$ fermions at each site. As in the case of complex fermions, we generate a variational wave function by considering the ground state of a free-fermion Hamiltonian. In this case, the mean-field decoupling yields \be
H_c=-\frac{i}{2}\sum_{\langle jl\rangle}\xi_{jl} \left(\mathbf{c}_{j}^{\dagger}\cdot   \mathbf{c}^{\phantom\dagger}_{l }- \mathbf{c}_{l }^{\dagger}  \cdot\mathbf{c}^{\phantom\dagger}_{j }\right).
\ee 
The values of  $\xi_{jl}$ are real numbers and must obey the relation $\xi_{jl}=-\xi_{lj}$.  Since the lattice is bipartite, we can choose that in every bond $\langle jl \rangle$ the site $j$ belongs to  an even sublattice and $l$ to an odd sublattice.  We then perform the gauge transformation   $\mathbf c_j=i\tilde{\mathbf c}_j$ and $\tilde{\mathbf c}_l=\tilde{\mathbf c}_l$ for all $j,l$. The mean-field Hamiltonian   becomes \be
H_c=-\frac12\sum_{\langle jl\rangle}\xi_{jl}  (\tilde{ \mathbf{c}}_{j}^{\dagger}\cdot   \tilde{\mathbf{c}}^{\phantom\dagger}_{l }+\text{h.c.}), \label{HMF1}
\ee
which is formally identical to the mean-field Hamiltonian for the $f$ fermions in Eq. (\ref{HMF}) if $\chi_{ij}\in \mathbb R$. As a consequence, the zero-flux and $\pi$-flux Ans\"atze for the Majorana fermion representation generate the same spectrum as the one shown in Fig. \ref{fig:Mean-field-dispersion}. 

However,   the enlarged Hilbert space in the Majorana fermion representation is different. Contrary to the   quarter filling condition for  complex $f$ fermions,    the average density of $c$ fermions is not constrained to  a specific value.  The mean-field ground state in this parton construction  is then obtained by filling up all the negative-energy states in Fig. \ref{fig:Mean-field-dispersion}. For the zero-flux state, the low-energy spectrum   has nodal lines like the ones in the exactly solvable Kitaev model on the hyperhoneycomb lattice \cite{Mandal2009,Lee2014,OBrien2016}. The nodal lines are illustrated in Fig. \ref{fig:nodalline} in Appendix \ref{sec:Details MFT}. The spectrum of the $\pi$-flux state in Fig. \ref{fig:pi-flux dispersion} also shows nodal lines; in this case we have observed numerically that  there are pairs of nodal lines connected by half of a reciprocal lattice vector of the base-centered orthorhombic lattice. Having identified the mean-field ground state, we obtain a trial wave function in the physical Hilbert space using a Gutzwiller projector $\mc P_c=\prod_{j}\left[\frac12+\frac12\prod_\gamma(1-2c^\dagger_{j\gamma}c^{\phantom\dagger}_{j\gamma})\right]$ to impose   the $\mathbb Z_2$ constraint in Eq. (\ref{Z2constraint}).

\subsection{Variational Monte Carlo results\label{VMC}}

To assess the viability of the proposed parton mean-field theories,
we now enforce the local constraints exactly by considering a Gutzwiller
projection of the mean-field wave functions \cite{Gros1989}. 

For complex fermions, we use the mean-field ground state from Eq. (\ref{HMF}). Both the zero-flux and $\pi$-flux states are considered, see Fig.
\ref{fig:ansatze}. To enforce the single-occupancy constraint in Eq. (\ref{U1constraint}), we generate physically allowed real-space configurations

\begin{equation}
\left|\left\{ j_{a}^{1}\right\} ,\,\left\{ j_{b}^{2}\right\} ,\,\left\{ j_{c}^{3}\right\} ,\,\left\{ j_{d}^{4}\right\} \right\rangle =\prod_{m=1}^{4}\prod_{\mathbf{r}_{m}=1}^{N/4}f_{m}^{\dagger}\left(\mathbf{r}_{m}\right)\left|\emptyset\right\rangle ,\label{eq:real_space_gutz}
\end{equation}
where $j_{a}^{m}$ denotes the position, at site $j$, of the $a$-th
fermion with color $m$. The overlap of \eqref{eq:real_space_gutz}
with the mean-field state is 
\begin{equation}
\Psi\left(\left\{ j_{a}^{1}\right\} ,\left\{ j_{b}^{2}\right\} ,\left\{ j_{c}^{3}\right\} ,\left\{ j_{d}^{4}\right\} \right)=\prod_{m=1}^{4}\Phi\left[\left\{ j^{m}\right\} \right].\label{eq:gutz_complex}
\end{equation}
Here, $\Phi\left[\left\{ j^{m}\right\} \right]$ is the Slater determinant
for one fermion species 
\begin{equation}
\Phi\left[\left\{ j^{m}\right\} \right]=\left|\begin{array}{cccc}
\zeta_{1}\left(j_{1}^{m}\right) & \zeta_{2}\left(j_{1}^{m}\right) & \cdots & \zeta_{N/4}\left(j_{1}^{m}\right)\\
\vdots & \vdots & \ddots & \vdots\\
\zeta_{1}\left(j_{N/4}^{m}\right) & \zeta_{2}\left(j_{N/4}^{m}\right) & \cdots & \zeta_{N/4}\left(j_{N/4}^{m}\right)
\end{array}\right|,\label{eq:slater_dirac}
\end{equation}
and $\zeta_{\nu}\left(j\right)$ is the amplitude of the fermion at
site $j$ in the $\nu$th eigenfunction of the mean-field Hamiltonian (\ref{HMF}):
$\zeta_{\nu}\left(j\right)\equiv\left\langle j|\nu\right\rangle $. 

We carry on variational Monte Carlo calculation using this wave function.
We describe the hyperhoneycomb lattice as a base-centered orthorhombic
lattice with an eight-point basis described in Appendix
\ref{sec:Hyperhoneycomb}, and thus the number of sites is
given by $N=8L^{3}$, with $L=3$, $4$, $5$, and $6$. We then randomly place each color at $N/4$ sites of our lattice.
Our Monte Carlo moves consists in exchanging a random pair of sites
containing distinct colors. We allow for moves involving sites far
away --- and which would not otherwise interact directly via the Hamiltonian
--- because this improves the sampling over the space of configurations.
We accept or reject these moves according to the general Metropolis
algorithm. The probability of accepting or rejecting each configuration
is proportional to the weight of the wave function
\begin{equation}
p_{\left\{ j\right\} }\propto\left|\prod_{m=1}^{4}\Phi\left[\left\{ j^{m}\right\} \right]\right|^{2}.\label{eq:prob}
\end{equation}
After $N_{{\rm exc}}$ of such exchanges attempts, we are said to
have performed a Monte Carlo sweep, and after every sweep, we compute
the ground state energy $E_{0}$. $N_{{\rm warm}}$ sweeps are performed
before measurements of physical quantities for ``thermalization''
while we consider $N_{{\rm mes}}$ measurements sweeps. We typically
use $N_{{\rm exc}}\sim10^{3}$ and $N_{{\rm warm}}=N_{{\rm mes}}\sim10^{5}$. 

For the Majorana fermion representation of the pseudospin and pseudo-orbital
operators, we consider the mean-field ground state from Eq. (\ref{HMF1}), already
written in terms of the three complex $c$ fermions. Again, both the
zero-flux and $\pi$-flux states are considered, see Fig. \ref{fig:ansatze}. A Gutzwiller
projection of these mean-field states imposes that a site can either
have no $c$ fermions, $\left|\emptyset\right\rangle $, or two $c$
fermions. For convenience, we follow Ref. \cite{Wang2009} and introduce
three states 
\begin{equation}
\left|X\right\rangle =c_{y}^{\dagger}c_{z}^{\dagger}\left|\emptyset\right\rangle ,\;\left|Y\right\rangle =c_{z}^{\dagger}c_{x}^{\dagger}\left|\emptyset\right\rangle ,\;\left|Z\right\rangle =c_{x}^{\dagger}c_{y}^{\dagger}\left|\emptyset\right\rangle .\label{eq:xyz_def}
\end{equation}
For any given configuration of these states, specified by the real-space location of the $\left|X\right\rangle $, $\left|Y\right\rangle $,
and $\left|Z\right\rangle $ states (at sites $\left\{ x_{i}\right\} $
, $\left\{ y_{j}\right\} $, and $\left\{ z_{m}\right\} $, respectively),
the projected wave function assigns an amplitude $\Psi\left(\left\{ x_{i}\right\} ,\left\{ y_{j}\right\} ,\left\{ z_{m}\right\} \right)$
to it. The locations of the $\left|\emptyset\right\rangle $ states
are automatically specified. Once we have constructed the mean-field
wave function, we generate a random initial state in which we distribute
each state, $\left|X\right\rangle $, $\left|Y\right\rangle $, $\left|Z\right\rangle $,
and $\left|\emptyset\right\rangle $, over $N/4$ distinct sites: $\left\{ x_{i}\right\} =\left\{ x_{1},x_{2},\ldots,x_{N/4}\right\} $,
etc. As in the case of complex fermions, our Monte Carlo moves consists
in exchanging random pair of sites containing distinct states and
the algorithm works in the same way. 

\begin{figure}
\begin{centering}
\includegraphics[width=.95\columnwidth]{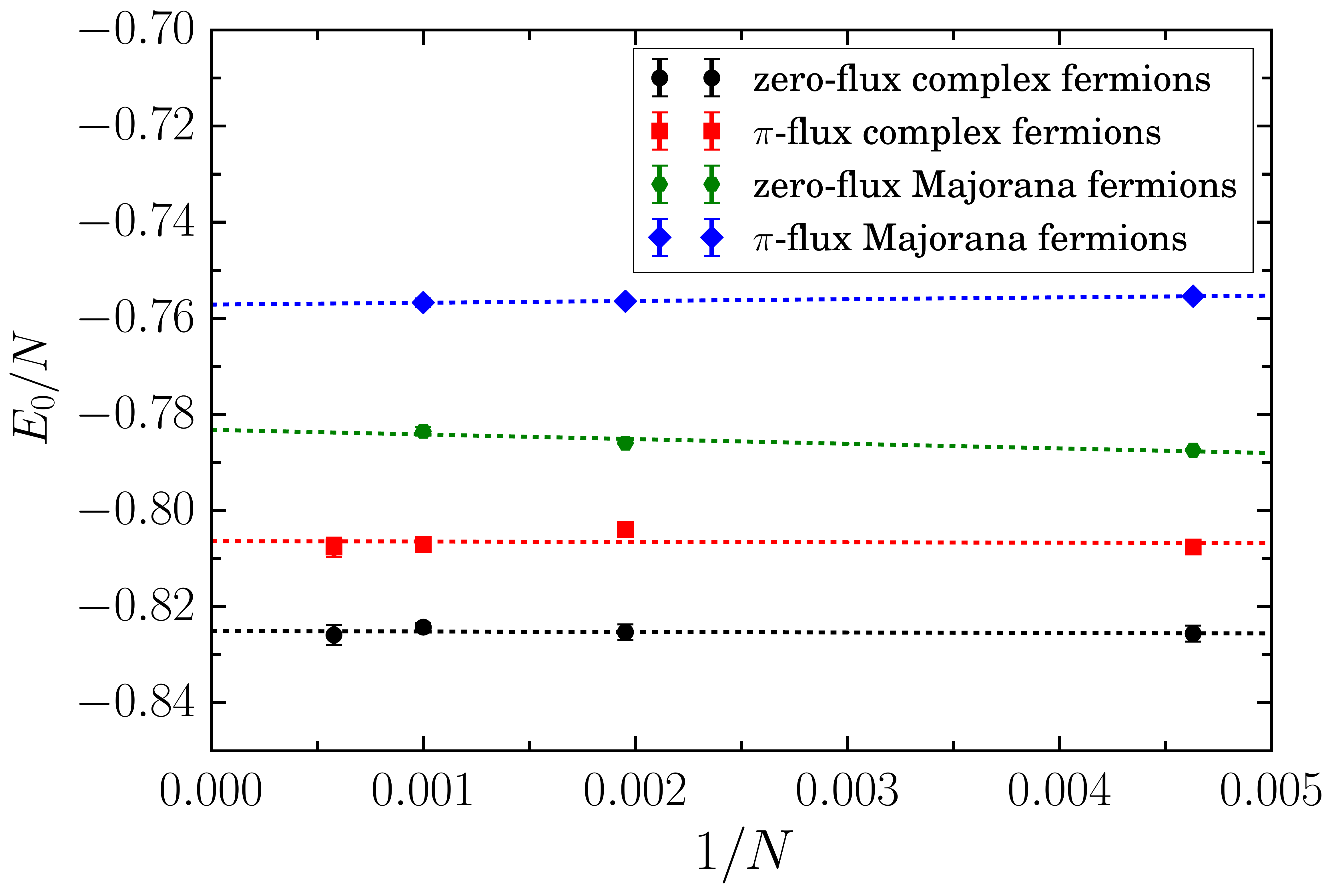}
\par\end{centering}

\caption{\label{fig:e0_vmc} (Color online) Variational Monte Carlo ground state energy, per site and in units
of $J$, for the different mean-field states as a function of the
inverse of the particle number. The dashed lines are linear extrapolations
to the data.}
\end{figure}

Figure \ref{fig:e0_vmc} shows the VMC results for the ground state energies of all four considered Ans\"atze at the different system sizes. As we can see, the results do not vary much with the system size and the extrapolated results for $N\rightarrow\infty$ are presented in Table \ref{tab:e0_vmc}. The ground state energies calculated at the mean-field level are also shown for comparison. As anticipated in Sec. \ref{meanfield}, the variational energies for the SU(4) model on the hyperhoneycomb lattice are comparable with those of the honeycomb lattice \cite{Corboz2012}, providing support to the feasibility of a spin-orbital liquid ground state. However, there are two significant differences: (i) the projected wave function with the lowest variational energy is the zero-flux state of complex fermions and (ii) the relative energy difference between our best variational state and the next candidate, the $\pi$-flux state for complex fermions, is roughly $2\%$, hinting at a fiercer
competition between the different variational states in the hyperhoneycomb lattice.

The sizable differences between the energies indicate that correlation effects beyond the parton mean-field theory are important to determine the most competitive ground state from our considered subset. This is equivalent to affirm that interactions between partons mediated by gauge fields are very important. However, we stress that strong interactions do not necessarily mean that the non-interacting trial state is \emph{qualitatively} incorrect. One procedure to determine the stability of a spin liquid against fluctuations assumes the mean-field states as starting points, integrates the high-energy fermions, and analyzes the effective action involving the gauge field and low-energy fermions. If the resulting perturbations to the non-interacting action are all irrelevant in the renormalization group sense, the spin liquid is stable \cite{Ran2007}. In the following, we will study the stability of this QSOL with an alternative approach using VMC. We will study the energetics of a modified wave function that incorporates the order parameter $\Delta$ of a symmetry-breaking phase \cite{Iaconis2018}. The QSOL will be regarded as stable against the formation of a certain kind of order if the minimum variational energy corresponds to the case in which $\Delta=0$.  

%This means that correlation effects beyond the parton mean-field theory are important to determine the most competitive ground %state from our considered subset. We stress that strong interactions do not necessarily mean that the non-interacting trial state %is \emph{qualitatively} incorrect (for instance, strongly interacting Fermi liquids, are Fermi liquids nonetheless).

%\textcolor{red}{[The referee made a comment that ``If [correlation effects are] quite
%important, the VMC method may not be appropriate to clarify the ground
%state in the system.'' Perhaps his point is that, if the energy changes so much, why should we believe that the projected wave function has anything to do with the original Fermi sea state? We could point out that such large differences between mean-field and VMC energies are also obtained on the honeycomb lattice studied by Corboz et al.. As far as I understand, the fact that interactions are strong does not mean that the state is qualitatively incorrect. Our problem could be analogous to strongly interacting Fermi liquids, which are Fermi liquids nonetheless. But I wouldn't trust the shape of the Fermi surface.]} 

\begin{table}
\caption{\label{tab:e0_vmc}Mean-field ($E_{\text{MF}}$) and  VMC ($E_0$) ground state energy, per site and in units of $J$, for the different
mean-field states.}
\begin{centering}
\begin{tabular}{|c|c|c|}
\hline 
Ansatz & $E_{\text{MF}}$&$E_{0}/N$\tabularnewline
\hline 
\hline 
Complex fermions zero-flux &$ 0.164 $&$-0.825\left(1\right)$\tabularnewline
\hline 
Complex fermions $\pi$-flux &$ 0.168 $&$-0.806\left(2\right)$\tabularnewline
\hline 
Majorana fermions zero-flux & $ -0.280 $&$-0.783\left(1\right)$\tabularnewline
\hline 
Majorana fermions $\pi$-flux & $ -0.221 $&$-0.757\left(1\right)$\tabularnewline
\hline 
\end{tabular}
\par\end{centering}

\end{table}

\begin{figure}
\begin{centering}
\includegraphics[width=.9\columnwidth]{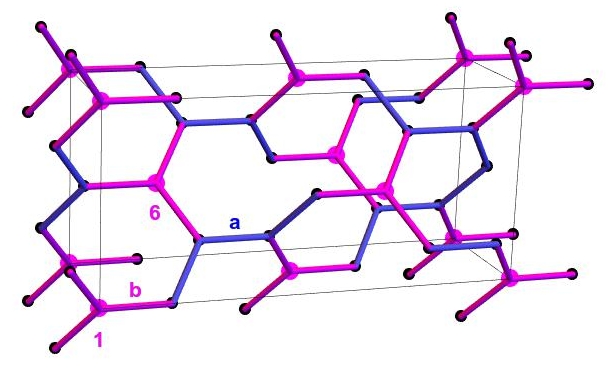}
\par\end{centering}
\caption{\label{fig:tetramer} (Color online) Graphical representation of a covering of tetramers on the 
hyperhoneycomb lattice. Following Eq. \ref{induce_tetramers}, the larger magenta disks indicate the sites in which the chemical potential is modified. Likewise, the hopping amplitudes are modified on the magenta bonds and leads to two types of non-equivalent bonds that are labeled $a$ and $b$. }
\end{figure}

We will now check the stability of this variational state against tetramerization \cite{Lajko2013}. In a tetramerized state, we observe the formation of four-site singlet plaquettes preserving the SU(4) symmetry but breaking the translational symmetry  \cite{Li1998,Bossche2001,Lajko2013}. A possible tetramer covering of the hyperhoneycomb lattice is illustrated in Fig. \ref{fig:tetramer}. The four-site plaquettes are centered on sites in sublattices $A_{1}$ and $A_{6}$. We have tested the stability of the zero-flux state against  this tetramerization pattern within VMC by considering variational wave functions generated by the mean-field Hamiltonian 
\be
H_f'=H_f +\sum_i\epsilon_i f^\dagger_if^{\phantom\dagger}_i, \label{induce_tetramers}
\ee
where $H_f$ is given by Eq. (\ref{HMF}) with modulated order parameters

\be
\chi_{ij}=\left\{\begin{array}{cl}
	t,& \text{if }i \in A_1 \cup   A_6 \text{ or }j \in A_1 \cup   A_6 ,\\
	1,&\text{otherwise}
	,\end{array}\right.
\ee
and $\epsilon_i$ are sublattice dependent on-site energies given by 
\be
\epsilon_i=\left\{\begin{array}{cl}
\epsilon,& i\in A_1 \cup  A_6\\
0,&\text{otherwise}.\end{array}\right.
\ee
Both $\epsilon$ and $t$ are variational parameters. For $\epsilon=0$ and $t=1$, the mean-field Hamiltonian reduces to the one in Eq. (\ref{HMF}) and we recover the symmetric zero-flux ansatz.  For $t> 1$ ($t<1$), the ground state of $H_f'$ corresponds to state with stronger (weaker) bonds inside the plaquettes represented in Fig. \ref{fig:tetramer}.  In the limit $t\to \infty$, we would obtain a product state of four-site singlets.

\begin{figure}
\begin{centering}
\includegraphics[width=0.95\columnwidth]{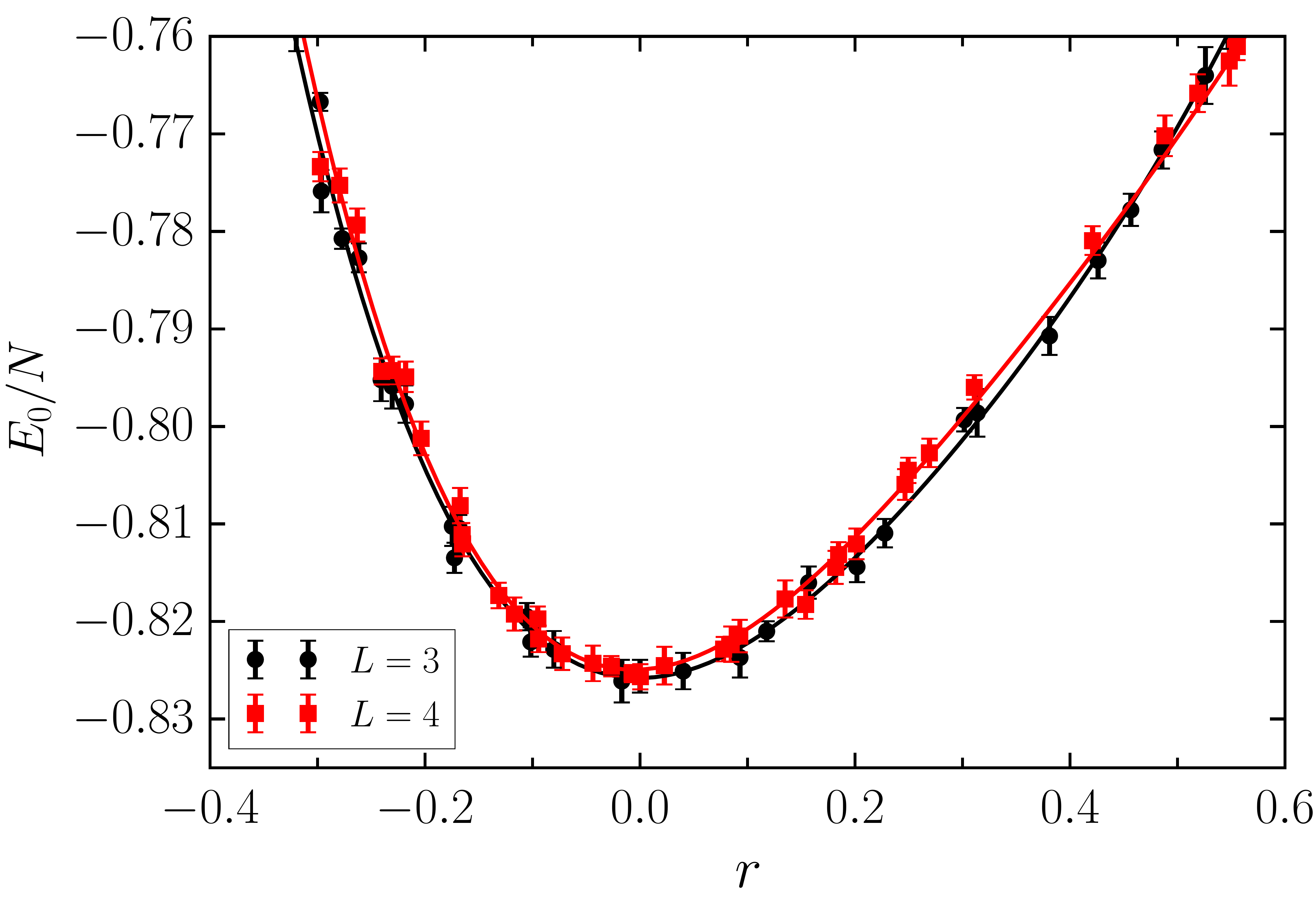}
\par\end{centering}
\caption{\label{fig:e0_tetra} (Color online) Variational Monte Carlo ground state energy for the zero-flux complex fermions ansatz as a function of the tetramerization  order parameter $r$ for two different system sizes. }
\end{figure}

We compute the energy of the projected wave functions as a function of $\epsilon$ and $t$ using VMC as described above for the spin-orbital liquid states. In order to quantify the degree of tetramerization of the wave functions, we first define the permutation operator on the links
\be
P_{ij}=\sum_{m,n=1}^4 S_m^n(i)S_n^m(j) \label{perm}.
%P_{ij}=\left(2\tilde{\mathbf s}_i\cdot \tilde{\mathbf s}_j+\frac12\right)\left(2\tilde{\boldsymbol \tau}_i\cdot \tilde{\boldsymbol \tau}_j+\frac12\right)
\ee 
Notice that Eq. (\ref{Hferm}) implies that $\bar H_{\text{eff}}=J \sum_{\left\langle ij \right\rangle} P_{ij}$. Beyond the mean-field level, the tetramerization order parameter is defined by \cite{Lajko2013}
\be
r= \frac{4}{5}\left(P_a-P_b\right)
\ee
where $P_a$ and $P_b$ are the ground state expectation values of Eq. (\ref{perm}) on the inequivalent bonds indicated by $a$ or $b$ in Fig. \ref{fig:tetramer}, respectively. The parameter $r=r(\epsilon,t)$ is normalized such that $r=1$  in the four-site plaquette product state. For each value of $\epsilon$, we select  the value of  $t=t_{\text{min}}(\epsilon)$ that gives the lowest   energy within VMC and compute the  corresponding tetramerization order parameter $r=r(\epsilon, t_{\text{min}}(\epsilon))$. Figure \ref{fig:e0_tetra}  shows the result for the energy as a function of $r$.  There is  little dependence on the system size $N=8L^3$ for $L=3$ compared to $L=4$. The lowest energy is obtained for $r=0$, from which we conclude that the zero-flux state is stable against tetramerization. The same conclusion was reached for the $\pi$-flux state on the honeycomb lattice \cite{Lajko2013}. 

Since the zero-flux state was stable against tetramerization, we now discuss its static spin-spin correlation function,
 a quantity which we can, in principle, calculate with our VMC approach through the average of the operator $P_{ij}-1/4$ \cite{Corboz2012}.  At the mean-field level, the correlation function of a three dimensional system with a Fermi surface displays the asymptotic behavior  $\langle M^z_i M^z_j \rangle\sim|\textbf{r}_{i}-\textbf{r}_{j}|^{-\alpha}$ with $\alpha=3$.
 Unfortunately, we were unable to verify corrections to $\alpha$ via VMC because the sizes of the system we are able to simulate are not large enough to accurately evaluate $\alpha$ (this limitation appears already in two-dimensional lattices \cite{Wang2009, Corboz2012}).

\section{Color-ordered states for finite Hund's coupling\label{sec:Hund}}
 
Although the results in Sec. \ref{meanfield} allow us to argue for a QSOL ground state on the SU(4) symmetric spin-orbital model, perturbations breaking the SU(4) symmetry can favor the onset of an ordered state. In this section, we investigate if the perturbations induced by nonzero Hund's coupling (Eq.(\ref{bigH2})) stabilize collinear spin-orbital orders on the hyperhoneycomb lattice through a combination of LFWT and VMC calculations.   

Let us consider product states of the form
\be
|\Psi\rangle =\bigotimes_i |\phi_i\rangle_i,
\ee
where $|\phi_i\rangle_i $ is an arbitrary  $j=3/2$ state at a site $i$ of the hyperhoneycomb lattice. Eq. (\ref{Hferm}) implies that the classical mean-field energy of these states at the SU(4)-symmetric model is $E=J \sum_{\langle ij\rangle}|\langle \phi_i|\phi_j\rangle|^2$. Therefore, the minimum classical energy for $|\Psi\rangle $ is $E=0$ and is obtained for any configuration in which the states of pairs of nearest-neighbor spins are orthogonal. This is achieved by taking $|\phi_i\rangle=|m_i\rangle$, where the set of colors $m_i=1,\dots,4$ specifies the classical configuration and $m_i\neq m_j$ when $i,j$ are nearest neighbors. Our study will be restricted to ordered states satisfying this condition. On the hyperhoneycomb lattice, the colors are assigned  according to the sublattice as follows:
\be
|\Psi(\{m_i\})\rangle =\bigotimes_{r=1}^8\bigotimes_{i\in A_r} |m_r\rangle_i.
\ee
Specifically, we will investigate the simplest ordered states, which are given by
\bea
\text{two-color}:& \{m_r\}=\{a,b,a,b,a,b,a,b\}\equiv \{a,b\},\label{two-color}\\
\text{four-color}:& \{m_r\}=\{a,b,c,d,b,a,d,c\}\nonumber\\&\equiv \{a,b,c,d\},\label{four-color}
\eea
where $a$, $b$, $c$ and $d$ are mutually distinct colors. The four-color state described by the color-ordering above is the analogue of the four-color state described on the honeycomb lattice in Ref. \cite{Corboz2012}.

\begin{figure}
\begin{centering}
\includegraphics[width=.9\columnwidth]{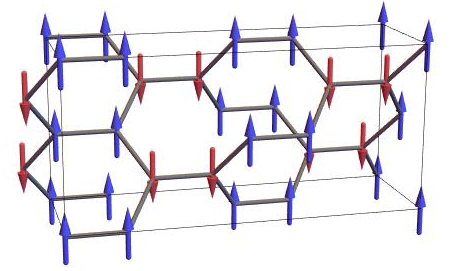}
\par\end{centering}

\caption{\label{fig:stripy_hyperhoneycomb} (Color online) Stripy phase with ordering in $M^{z}=\pm 3/2$. This is the only collinear ordered state stable at the linear flavor wave theory level among the two- and four-color ordered states.}
\end{figure}

Although the ordered states are conveniently written in the rotated frame, their physical interpretation requires their translation into the original pseudospin and pseudo-orbital quantum numbers. For example, notice that the two-sublattice transformation on the pseudo-orbitals in Eq. (\ref{eq:two-sublattice transformation}) implies that the state $\{a,b\}$ is not equivalent to $\{b,a\}$. This point is illustrated by 
\be
\{m_r\}=\{1,3\}, \quad \{m_r\}=\{2,4\}. \label{stripy}
\ee
These states represent a ferromagnetic order on $\tilde{s}^{z}$ and a N\'eel order on $\tilde{\tau}^{z}$. Applying the transformations given by Eq. (\ref{eq:Klein transformation}) and Eq. (\ref{eq:two-sublattice transformation}), these states correspond to a stripy order on the pseudospins $s^{z}$ \cite{Lee2014} and a ferromagnetic order on the pseudo-orbitals, with $\tau^{z}=+1/2$. In terms of the dipoles given by Eq. (\ref{eq:dipole}), $\{m_r\}=\{1,3\}$ and $\{m_r\}=\{2,4\}$ correspond to a stripy order with $M^{z}=\pm 3/2$, as represented in Fig. \ref{fig:stripy_hyperhoneycomb}. By contrast, the states $\{m_r\}=\{3,1\}$ and $\{m_r\}=\{4,2\}$ correspond to stripy ordered states of $M^{z}=\pm 1/2$ dipoles. 

We  treat quantum fluctuations  on top of the color-ordered states using a Holstein-Primakoff transformation for the generators of SU(4) \cite{Joshi1999}. The explicit forms of $\tilde{s}^a,\tilde{\tau}^b,\tilde{s}^a\tilde{\tau}^b$ in terms of these generators are given in Appendix \ref{sec:LFWT_operators} and are used to rewrite Eq. (\ref{bigH2}). Next, at each site $i$ in a given sublattice $A_r$,  with classical state $m_r$,  we define three flavors of bosons $b_{irn}$, $n\in \{1,\dots,4\} \setminus \{m_r\}$, which obey canonical commutation relations $[b^{\phantom\dagger}_{irn},b^\dagger_{jr'n'}]=\delta_{ij}\delta_{rr'}\delta_{nn'}$. The local operators are given by 
\bea
\hspace{-0.5cm}S^{m_r}_{m_r}(i)&=&1-\sum_{n\neq m_r}b^\dagger_{irn}b^{\phantom\dagger}_{irn},\\
\hspace{-0.5cm}S^{m_r}_{n}(i)&=&b^\dagger_{irn}\sqrt{1-\sum_{l\neq m_r}b^\dagger_{irl}b^{\phantom\dagger}_{irl}},\qquad n\neq m_r,\\
\hspace{-0.5cm}S^{l}_n(i)&=&b^\dagger_{irn}b^{\phantom\dagger}_{irl},\qquad l,n\neq m_r. 
\eea
With three bosons per site and eight sublattices, we have in total $24$ flavors of bosons.  

Within LFWT, we substitute the Holstein-Primakoff transformation into Hamiltonian (\ref{bigH2}) and expand the latter to quadratic order in the bosonic operators. After a Fourier transform to momentum space, the LFWT Hamiltonian can be cast in the form 

\bea
H_{fw}&=&\sum_{\mathbf k}\left(B^\dagger_{\mathbf k},B_{-\mathbf k}\right)\left(\begin{array}{cc}\mc H_{11}(\mathbf k)&\mc H_{12}(\mathbf k)\\
\mc H^\dagger_{12}(\mathbf k)&\mc H_{22}(\mathbf k)
\end{array}\right)\left(\begin{array}{c}B^{\phantom\dagger}_{\mathbf k}\\
B^{ \dagger}_{-\mathbf k}\end{array}\right)\nonumber\\
&&-\frac{3}{2}N\left(J_{a}+J_{b}+5J_{c}\right), \label{Hfw}
\eea
where   $B_{\mathbf k}$ is the 24-component vector of boson annihilation operators and $\mc H_{ab}$, with $a,b=1,2$, are $24\times 24$ matrices. Finally, the Hamiltonian is diagonalized by a Bogoliubov transformation and we obtain
\bea
H_{fw}&=&-\frac{3}{2}N\left(J_{a}+J_{b}+5J_{c}\right)+\underset{\textbf{k}}{\sum}\underset{\lambda=1}{\overset{24}\sum}\Omega_{\lambda}(\textbf{k})\nonumber\\
&&+\underset{\lambda=1}{\overset{24}\sum}\sum_{\mathbf k}\Omega_\lambda(\mathbf k)\left(\Xi^\dagger_{\mathbf k\lambda}\Xi^{\phantom\dagger}_{\mathbf k\lambda}+\Phi^\dagger_{\mathbf k\lambda}\Phi^{\phantom\dagger}_{\mathbf k\lambda}\right), \label{Hfw_zero_energy}
\eea
where $\Omega_\lambda(\mathbf k)$ are the dispersion relations of the ``magnons'' created by the operators $\Xi^\dagger_{\mathbf k\lambda}$ and $\Phi^\dagger_{\mathbf k\lambda}$. It is important to point out that real values for $\Omega_\lambda(\mathbf k)$ can be ensured only if the classical state corresponds to a local minima of the mean-field theory. Henceforth, only ordered states satisfying this condition will be regarded as stable. 

The four-color states given by Eq. (\ref{four-color}) were found to be unstable at the LFWT approximation, since infinitesimal values of Hund's coupling generates imaginary frequencies in the dispersion $\Omega_\lambda(\mathbf k)$. Indications of this instability appear already at the SU(4)-symmetric point, where they display zero-energy flat bands $\Omega_\lambda(\mathbf k)=0$ for all $\lambda$ and $\mathbf k$ that lead to the zero-point energy $E=-1.5NJ$. This remarkably low energy is achieved because $H_{fw}$ is the sum of two-site disconnected clusters throughout the lattice \cite{Corboz2011,Corboz2012}. The zero-point fluctuations are then minimized, ensuring an energy gain of $-J$ per bond. Such characteristic of $H_{fw}$ also implies that the four-colored states on the hyperhoneycomb lattice displays a degeneracy analogous to the one observed on the honeycomb lattice \cite{Corboz2012}. This indicates the absence of lattice symmetry breaking and contradicts the formation of an ordered state \cite{Corboz2012}. Hence, four-colored states are not good candidates for the ground state of Eq. (\ref{bigH2}).

\begin{figure}

\begin{centering}
\subfloat[\label{fig:eta=0_dispersion}]{\begin{centering}
\includegraphics[width=.9\columnwidth]{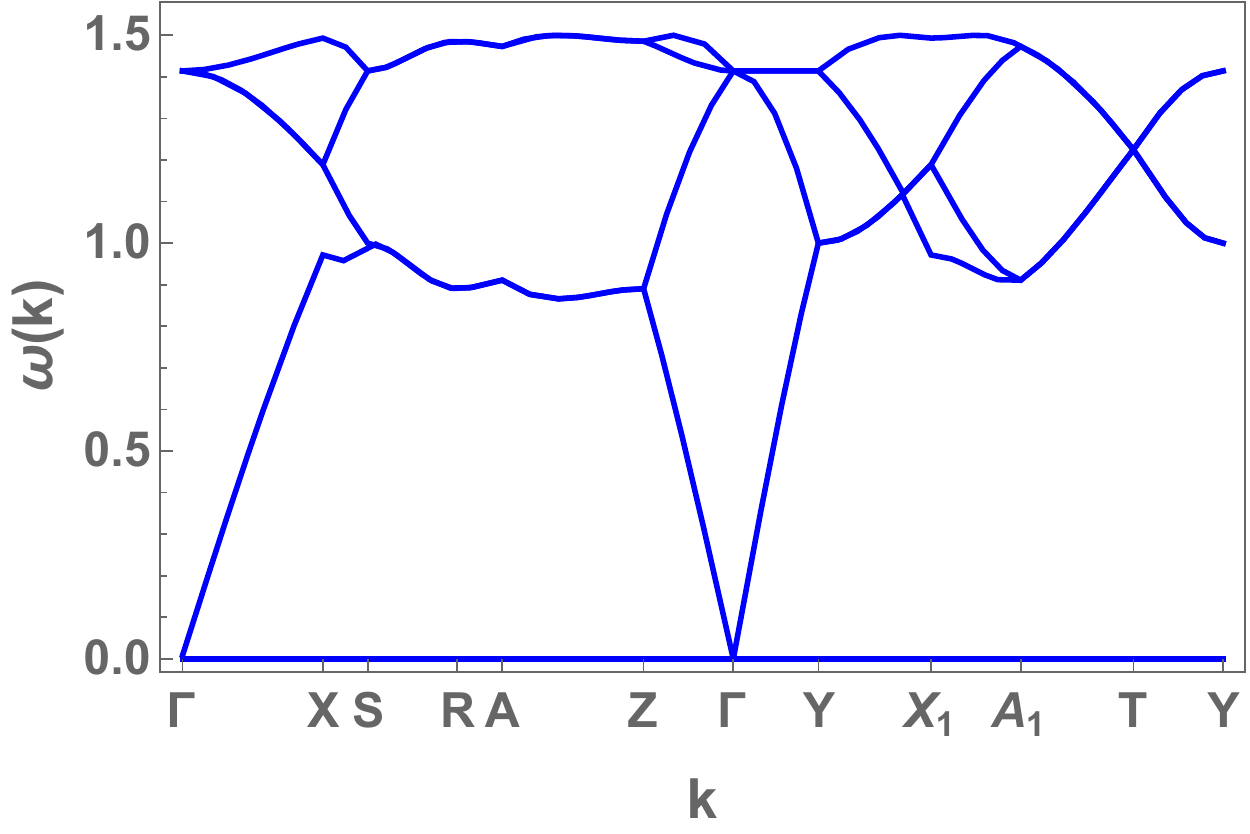}
\par\end{centering}}\par\end{centering}

\begin{centering}
\subfloat[\label{fig:eta=0.1_dispersion}]{\begin{centering}
\includegraphics[width=.9\columnwidth]{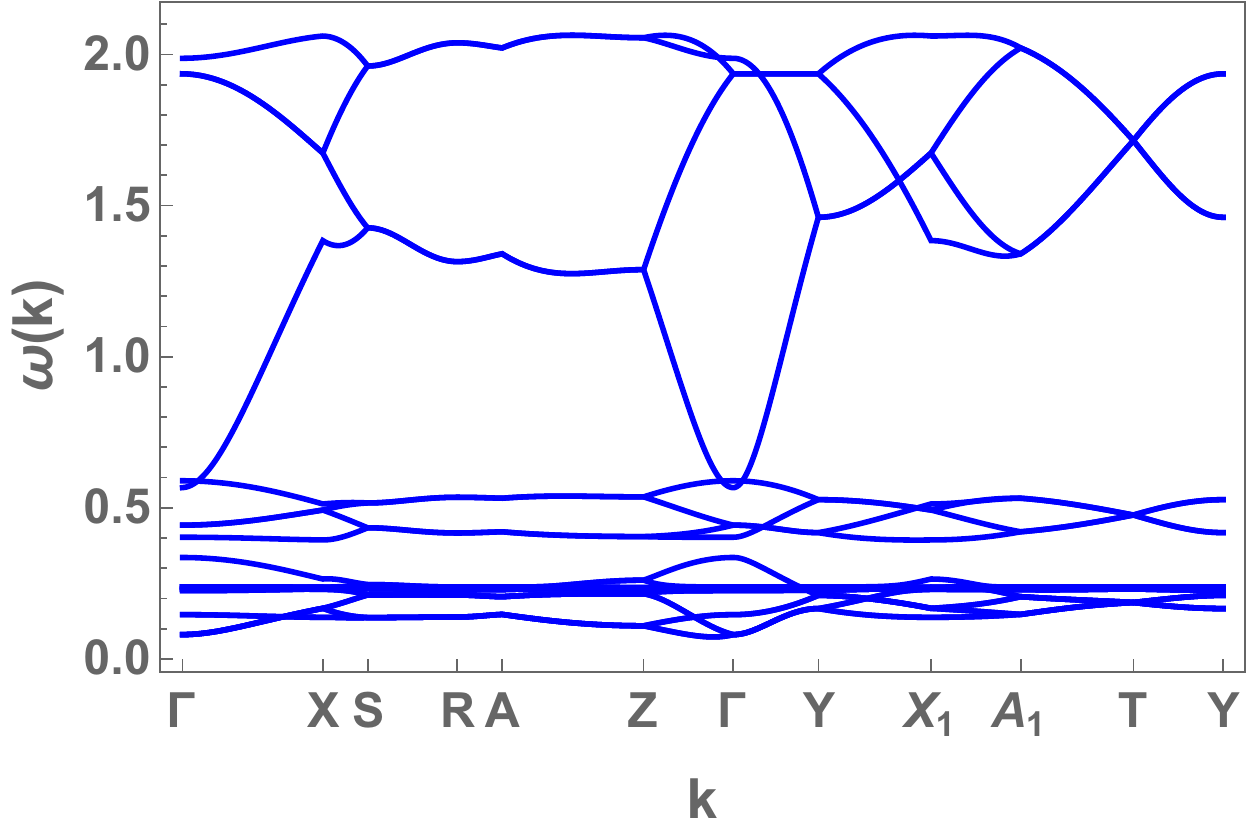}
\par\end{centering}}
\par\end{centering}

\caption{\label{fig:LFWT-dispersion}  Linear flavor wave dispersion of the $M^{z}=\pm 3/2$ stripy phase for (a) $\eta=J_{H}/U=0$ and (b) $\eta=J_{H}/U=0.1$. Figure (a) shows the spectra of eight bands, sixteen degenerate bands with zero energy and Goldstone modes at the $\Gamma$ point. Figure (b) shows the dispersion of the 24 bands after the inclusion of Hund's coupling induced perturbations. The lack of Goldstone modes is due to the absence of continuous symmetry on the underlying Hamiltonian.}
\end{figure}

Let us now consider the LFWT approximation of the two-color ordered states starting from the SU(4)-symmetric point. Using the Holstein-Primakoff transformation on Eq. (\ref{HSU4temp}), the LFWT Hamiltonian will be determined by \cite{Corboz2012}  
\be
\mathcal{H}_{\text{SU(4)},ij}\rightarrow Z_{ij,m_{s(i)},m_{s(j)}}^{\dagger}Z_{ij,m_{s(i)},m_{s(j)}}-1,
\ee
where $Z_{ij,m_{s(i)},m_{s(j)}}=b_{j,s(j),m_{s(i)}}+b_{i,s(i),m_{s(j)}}^{\dagger}$ with $s(x)$ being the sublattice index of the site $x$. If we first define
\bea
p_{1}(\mathbf{k}) & =&2\cos(k_{x}-k_{y})\cos(2k_{z}),\nonumber \\
p_{2}(\mathbf{k}) & =&5-\cos\left[2(k_{x}-k_{y})\right]+2\cos\left[2(2k_{x}+k_{y})\right]\nonumber \\
    && +2\cos\left[2(k_{x}+2k_{y})\right]-2\cos(4k_{z})\sin^{2}(k_{x}-k_{y}),\nonumber \\
p_{3}(\mathbf{k}) & =&8\cos(2k_{x}+k_{y})\cos(k_{x}+2k_{y})\cos(2k_{z}), \label{eq:pi}
\eea 
the analytical expressions for the flavor-wave dispersion in this case are
\bea
\epsilon_{1}(\mathbf{k})& =&\frac{J}{2}\sqrt{6-p_{1}(\mathbf{k})-\sqrt{p_{2}(\mathbf{k})+p_{3}(\mathbf{k})}},\nonumber\\
\epsilon_{2}(\mathbf{k})& =&\frac{J}{2}\sqrt{6-p_{1}(\mathbf{k})+\sqrt{p_{2}(\mathbf{k})+p_{3}(\mathbf{k})}},\nonumber\\
\epsilon_{3}(\mathbf{k})& =&\frac{J}{2}\sqrt{6+p_{1}(\mathbf{k})-\sqrt{p_{2}(\mathbf{k})-p_{3}(\mathbf{k})}},\nonumber\\
\epsilon_{4}(\mathbf{k})& =&\frac{J}{2}\sqrt{6+p_{1}(\mathbf{k})+\sqrt{p_{2}(\mathbf{k})-p_{3}(\mathbf{k})}}.\label{eq:SU4LFWTdisp}
\eea
The bands above are twofold degenerate and are displayed in Fig. \ref{fig:eta=0_dispersion}(a). The spectrum also presents 16 degenerate flat bands with zero energy, which will be explained below.  

The similar behavior of all two-color states on the SU(4) symmetric point contrasts with the different ways they are affected by finite Hund's coupling perturbations. For  $\eta=J_{H}/U>0$, only the $M^{z}=3/2$ stripy states display real and non-negative energies $\Omega_\lambda(\mathbf k)$ for all $\lambda$ and $\mathbf k$ up to $\eta \approx 0.125$. Finite Hund's coupling also includes flavors on $H_{fw}$ that were not explicitly present at the SU(4)-symmetric point. For example, the $\mathcal{H}_{b,ij}$ given by Eq. (\ref{Hbtemp}) for the $\{1,3\}$ state gives rise to

\bea
\mathcal{H}_{b,ij}& \rightarrow& Z_{ij,1,3}^{\dagger}Z_{ij,1,3}-1 \nonumber \\
&&+b_{i,s(i),4}^{\dagger}b_{i,s(i),4}+b_{j,s(j),2}^{\dagger}b_{j,s(j),2}\nonumber \\
&&-b_{i,s(i),4}^{\dagger}b_{j,s(j),2}-b_{j,s(j),2}^{\dagger}b_{i,s(i),4}.
\eea
The LFWT Hamiltonian of $\mathcal{H}_{c,ij}$ (Eq. (\ref{Hctemp})) also leads to the inclusion of other flavors and ensures that the eigenstates of the complete Hamiltonian are bogolons. Figure \ref{fig:eta=0.1_dispersion} shows the 24 flavor-wave bands of the stripy state for $\eta=0.1$, in which the 16 lower-energy bands are non-degenerate. In contrast to the SU(4)-symmetric case, there is no Goldstone boson at $\Gamma$ for $\eta>0$, since all continuous symmetries are explicitly broken by $\mathcal{H}_{b,ij}$ and $\mathcal{H}_{c,ij}$. The widths of the lower-energy bands vanish in the limit $\eta \rightarrow 0^{+}$ and all these bands become degenerate at $\omega=0$, providing an explanation for the flat bands shown in Figure \ref{fig:eta=0_dispersion}.

\begin{figure}

\begin{centering}
\includegraphics[width=0.95\columnwidth]{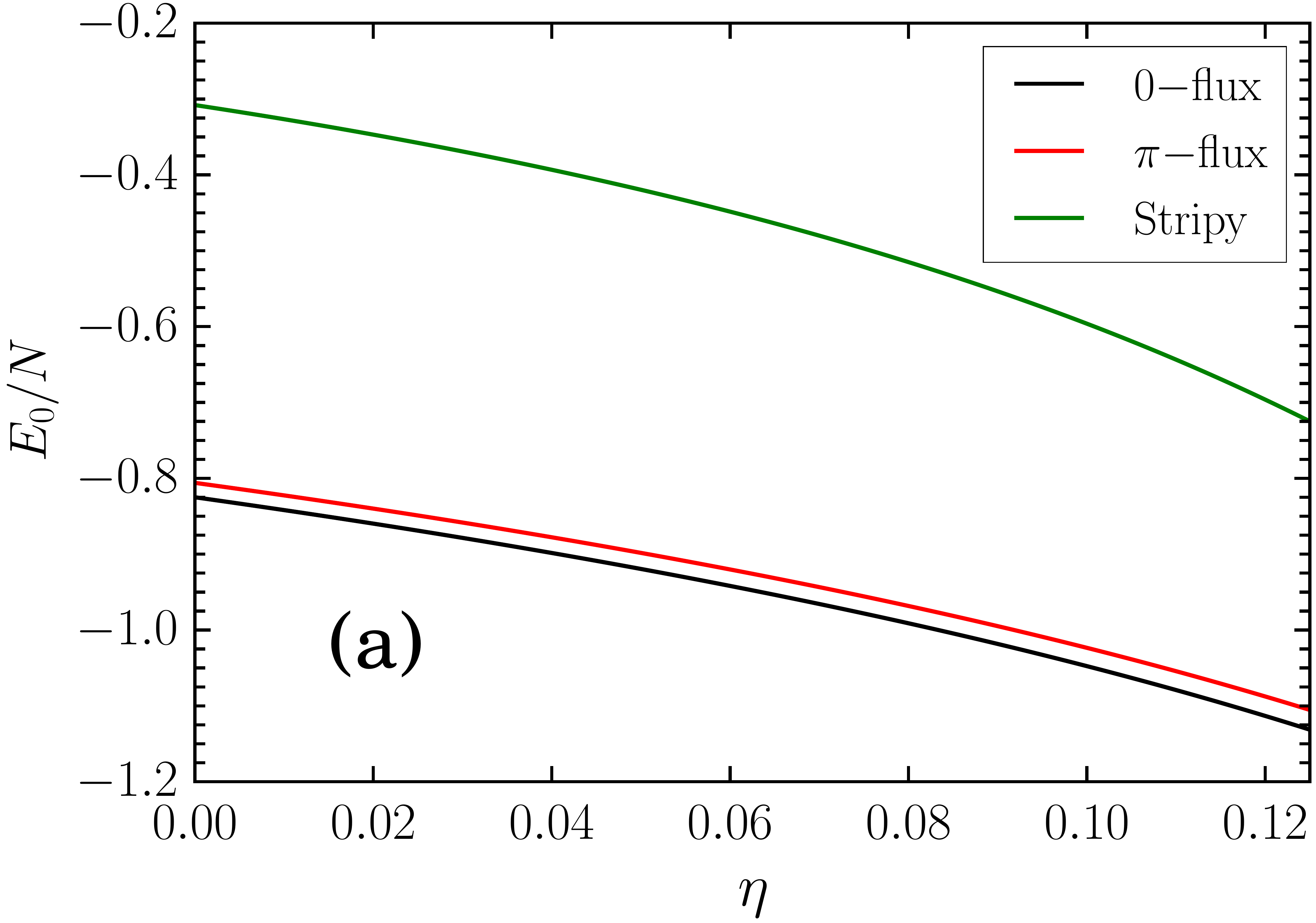}
\par\end{centering}

\begin{centering}
\includegraphics[width=0.95\columnwidth]{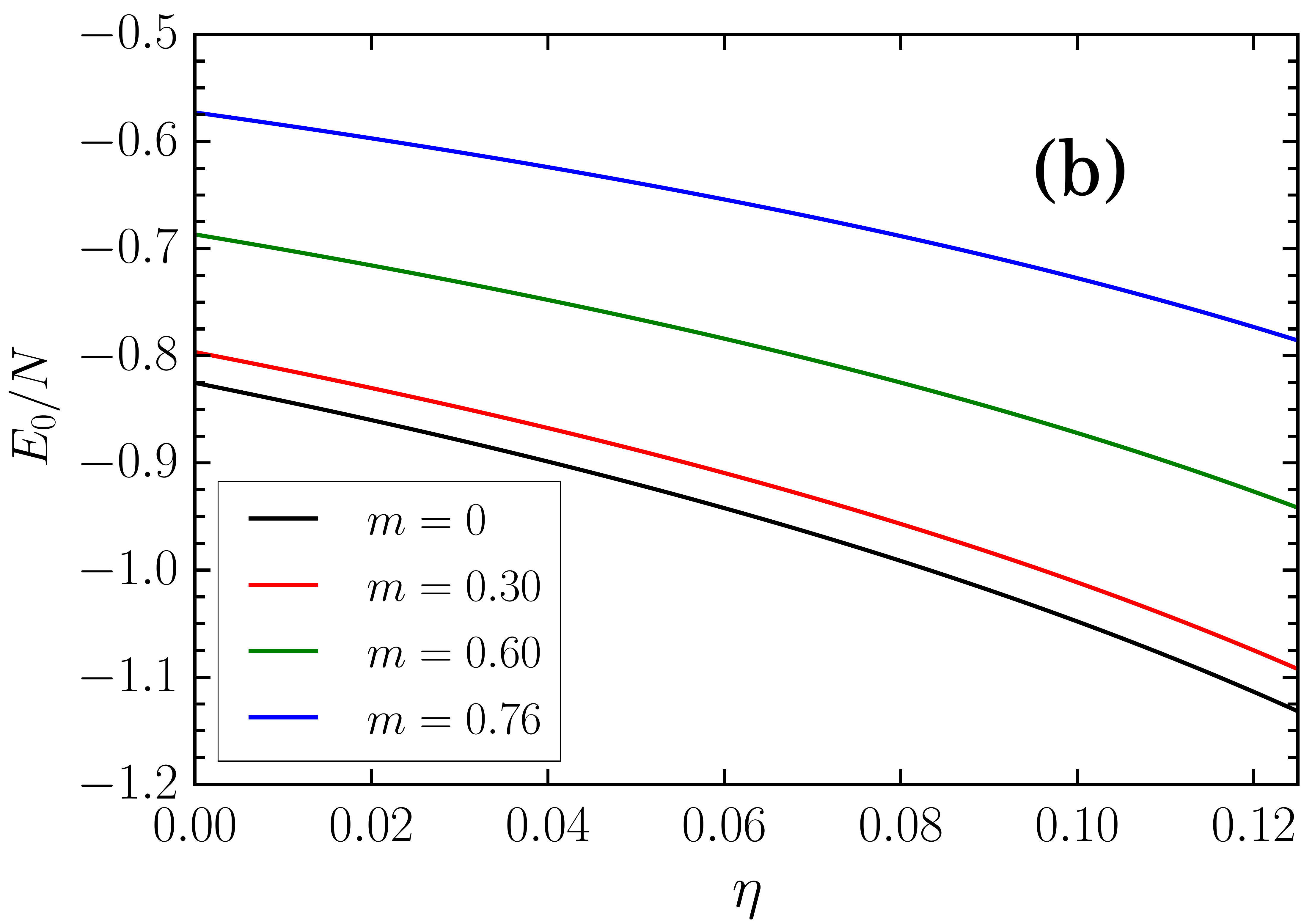}
\par\end{centering}

\caption{\label{fig:LFWTenergy} (Color online) (a) Energy per site of three different states as a function of the ratio $\eta=J_{H}/U$.The energy of the stripy phase was estimated using linear flavor wave theory, whereas the energy of the QSOLs based on complex fermions was evaluated with VMC. (b) VMC ground state energy for zero-flux complex fermions in the presence of a staggered potential favoring stripy order. The four curves correspond to different values of the local moment $m$.}
\end{figure}

The zero-point energy of the stripy state at the LFWT level is displayed in Fig. \ref{fig:LFWTenergy}(a). The energy of the stripy state at the SU(4) symmetric point is $E_{\text{stripy}}\approx-0.3079NJ$, close to the value of $E\approx-0.314NJ$ found for the two-color ordered states on the honeycomb lattice \cite{FranciscoKim2017}. Although this energy is not variational, it is significantly higher than the ones of the previously studied QSOLs and indicates that the stripy phase is not competitive at this point. Nevertheless, the zero-point energy decreases with increasing values of Hund's coupling. This prompted us to calculate the energy of the complex fermion QSOLs for the perturbed Hamiltonian in Eq. (\ref{bigH2}) using VMC. These results are also displayed in Fig. \ref{fig:LFWTenergy}(a) and indicate that the QSOL states remain energetically favored even in the perturbed model. 

Variational results in favor of the QSOL stability against the formation of the stripy order were also found using VMC. To include magnetic orders in our variational scheme, we add a color dependent local site
energy $\tilde{\varepsilon}_{im}$ to the mean-field Hamiltonian in
Eq. \ref{HMF}, which define our trial states:
\begin{equation}
\mathcal{H}_{f}^{\prime\prime}=\mathcal{H}_{f}-\sum_{i}\sum_{m=1}^{4}\tilde{\varepsilon}_{im}f_{im}^{\dagger}f_{im}.\label{eq:induce_stripy}
\end{equation}
In particular, we consider the stripy order listed in Eq. (\ref{two-color}) on
top of the complex fermions zero-flux Ansatz. We do so by setting
$\tilde{\varepsilon}_{i1(3)}=\tilde{\varepsilon}$ in sublattices
$A_{{\rm odd}}$  $\left(A_{{\rm even}}\right)$, with $\tilde{\varepsilon}_{i1(3)}=0$ otherwise. The
local moment associate to this order is given by
\begin{equation}
m=\frac{4}{N}\left[\sum_{i\in A_{\rm{odd}}}\left\langle f_{i1}^{\dagger}f_{i1}\right\rangle +\sum_{i\in A_{\rm{even}}}\left\langle f_{i3}^{\dagger}f_{i3}\right\rangle \right]-1.\label{eq:m_stripy}
\end{equation}
We then have that $m=0$ for $\tilde{\varepsilon}=0$ and $m\rightarrow1$
as $\tilde{\varepsilon}\rightarrow\infty$. In Fig. \ref{fig:LFWTenergy}(b) we show the resulting
ground state energy for different values of $m$ as a function of
$\eta$. We clearly see that states with $m>0$, displaying stripy order, have higher energy than the SU(4)-symmetric QSOL. We thus confirm the LFWT results showing that this spin-liquid state is not unstable towards collinear ordering for any value of $\eta$.

\section{Conclusions\label{sec:Conclusions}}

We have derived  an effective model for $4d^1$ and $5d^1$ Mott insulators in tricoordinated lattices in the limit of strong spin-orbit coupling. For vanishing Hund's coupling, the model for $j=3/2$ local moments has an SU(4) symmetry which can be made explicit using a Klein transformation. We then used fermionic parton mean-field theories to propose quantum spin-orbital liquid states on the hyperhoneycomb lattice. Variational Monte Carlo simulations showed that the lowest-energy trial wave function is a Fermi sea of complex fermions at quarter filling with zero gauge flux through every plaquette. In contrast with the nodal-line spectrum of the  Kitaev model on the hyperhoneycomb lattice,  the zero-flux state of complex fermions has a large Fermi surface. We could verify that this does not translate into instability against tetramerization. The simplest ordered states were studied in the minimal model for arbitrary values of Hund's coupling through a combination of LFWT and VMC and we could confirm that they are energetically uncompetitive. Our present results do not indicate a transition from a spin-orbital liquid state to an ordered one through the studied perturbations.  

In the $j=1/2$ material $\beta$-Li$_2$IrO$_3$, sizeable Heisenberg exchange interactions move the system away from the Kitaev spin liquid phase and lead to incommensurate noncoplanar magnetic order \cite{BiffinB2014}. Ab initio studies on this compound indicated that interactions driven by other hopping mechanisms, longer-range interactions and slight distortions are essential to understand its ground state \cite{Katukuri2016}. One important open question is if such perturbations to the SU(4)-symmetric model would also appear on the hypothetical $j=3/2$ counterpart of this iridate and induce an analogue incommensurate spin-orbital order.

Ref. \cite{Yamada2018} mentions that 4/5$d^{1}$ materials could be synthesized from an oxide A$_{2}$MO$_{3}$ (M=Nb, Ta) or in the Zr- and Hf-based metal-organic frameworks. If such compound were synthesized with the hyperhoneycomb lattice structure, one could look for signatures of the zero-flux spin-orbital liquid in the temperature dependence of the magnetic specific heat $C(T)$. The prediction for a Fermi surface of fermionic partons coupled to a U(1) gauge field is $C(T)/T\sim -\ln T$ at low temperatures \cite{Holstein1973,Reizer1989}. This differs significantly from the prediction for the Kitaev spin liquid, in which $C(T)/T$ vanishes linearly with temperature \cite{Lee2014}.

\acknowledgements
We thank V. S. de Carvalho and E. Miranda for helpful discussions. This work was supported
by the Brazilian agencies FAPESP (W.M.H.N., E.C.A.) and CNPq (R.G.P., E.C.A.).

\appendix

\section{The Ideal Hyperhoneycomb Lattice\label{sec:Hyperhoneycomb}}

Here we present the description of  the hyperhoneycomb lattice 
as a base-centered orthorhombic lattice with an eight-point
basis. The   position of the basis is given by
\begin{align}
\text{M}_{1}=(0,0,0), &  & \text{M}_{2}=(1,1,0), &  & \text{M}_{3}=(1,2,1), \nonumber\\
\text{M}_{4}=(2,3,1),&  & \text{M}_{5}=(3,3,2), &  & \text{M}_{6}=(4,4,2), \nonumber\\
 \text{M}_{7}=(4,5,3), &  & \text{M}_{8}=(5,6,3).\label{eq:basis points}
\end{align}
We consider the following primitive lattice vectors of the base-centered
orthorhombic lattice
\begin{equation}
\mathbf{a}_{1}=(2,4,0),\,\mathbf{a}_{2}=(-2,2,0),\,\mathbf{a}_{3}=(0,0,4).\label{eq:primitive lattice vectors}
\end{equation}
The corresponding reciprocal lattice vectors are
\begin{equation}
\mathbf{b}_{1}=\left(\frac{\pi}{3},\frac{\pi}{3},0\right),\,\mathbf{b}_{2}=\left(-\frac{2\pi}{3},\frac{\pi}{3},0\right),\,\mathbf{b}_{3}=\left(0,0,\frac{\pi}{2}\right).
\end{equation}
The high-symmetry points in the first Brillouin zone are given by\begin{align}
&\Gamma=(0,0,0), \;  \text{X}_{1}=\left(-\frac{7\pi}{18},\frac{\pi}{18},0\right), \;\text{Y}=\left(\frac{\pi}{6},\frac{\pi}{6},0\right),\nonumber\\
&\text{T}=\left(\frac{\pi}{6},\frac{\pi}{6},\frac{\pi}{4}\right),  \; \text{A}_{1}=\left(-\frac{7\pi}{18},\frac{\pi}{18},\frac{\pi}{4}\right),\nonumber\\
&\text{Z}=\left(0,0,\frac{\pi}{4}\right),  \;  \text{S}=\left(-\frac{\pi}{6},\frac{\pi}{3},0\right),\; \text{X}=\left(-\frac{5\pi}{18},\frac{5\pi}{18},0\right),  \nonumber\\
& \text{A}=\left(-\frac{5\pi}{18},\frac{5\pi}{18},\frac{\pi}{4}\right),\;\text{R}=\left(-\frac{\pi}{6},\frac{\pi}{3},\frac{\pi}{4}\right).
\end{align}

There are four distinct ten-site elementary loops on the hyperhoneycomb
lattice. In terms of the basis points defined in Eq. (\ref{eq:basis points}),
the loops are (see Fig. \ref{fig:hyperhoneycomb lattice}):
\begin{align}
P_{1} :&\; 1\rightarrow2\rightarrow3\rightarrow4\rightarrow5\rightarrow8\nonumber \\
 & \quad\rightarrow7\rightarrow6\rightarrow5\rightarrow4\rightarrow1,\nonumber\\
P_{2} :&\; 1\rightarrow2\rightarrow3\rightarrow6\rightarrow5\rightarrow8\nonumber \\
 & \quad\rightarrow7\rightarrow6\rightarrow3\rightarrow4\rightarrow1,\nonumber\\
P_{3} :& \;1\rightarrow2\rightarrow7\rightarrow6\rightarrow5\rightarrow4\nonumber \\
 & \quad\rightarrow3\rightarrow6\rightarrow7\rightarrow8\rightarrow1,\nonumber\\
P_{4} :& \;1\rightarrow2\rightarrow7\rightarrow8\rightarrow5\rightarrow4\nonumber \\
 & \quad\rightarrow3\rightarrow6\rightarrow5\rightarrow8\rightarrow1.
\end{align}
This  can be used to check that the ansatz in  Fig. \ref{fig:pi flux Majorana}
has   gauge flux $\Phi=\pi$ through all loops.

\section{Explicit Form of the Matrices Generating the Trial Wave Functions\label{sec:Details MFT}}

After fixing the bond variables $\chi_{ij}$ in the zero-flux or $\pi$-flux state, we can diagonalize the mean-field Hamiltonian in Eq. (\ref{HMF}) using Fourier transform. Since the four colors are decoupled at the mean-field level, here we drop the index $m=1,\dots,4$. The Hamiltonian for each color has the form\be
H_f=\sum_{\mathbf k} \sum_{r,r'=1}^8f^\dagger_{\mathbf k r}[\mc H^{\Phi}(\mathbf k)]_{rr'}f^{\phantom\dagger}_{\mathbf kr'},
\ee
where $\mathbf k$ is a vector in the first Brillouin zone of the base-centered orthorhombic lattice, $r,r'$ are the sublattice indices, and $\mc H^{\Phi}(\mathbf k)$ are $8\times 8$ matrices labeled  by the  uniform gauge flux $\Phi=0,\pi$. Here we use the notation 
\be
\Lambda^{abc}=\sigma^a\otimes\sigma^b\otimes \sigma^c,
\ee
where $a,b,c\in \{0,1,2,3\}$ with $\sigma^0=\mathbb I_{2\times 2}$ the   identity matrix and $\sigma^{1,2,3}=\sigma^{x,y,z}$ the Pauli matrices. The Hamiltonian matrices for the zero-flux and $\pi$-flux states are given respectively by

\begin{align}
\mc H^{0}(\mathbf k) & =-\cos(k_{x}+k_{y})\Lambda^{001}-\sin(k_{x}+k_{y})\Lambda^{002}\nonumber \\
 & \quad+\cos k_{z}\left(-\cos k_{y}\Lambda^{011}+\sin k_{y}\Lambda^{012}\right)\nonumber \\
 & \quad+\sin k_{z}\left(-\cos k_{y}\Lambda^{021}+\sin k_{y}\Lambda^{022}\right)\nonumber \\
 & \quad+\cos k_{z}\left(-\cos k_{x}\Lambda^{111}+\sin k_{x}\Lambda^{112}\right)\nonumber \\
 & \quad-\sin k_{z}\left(-\cos k_{x}\Lambda^{121}+\sin k_{x}\Lambda^{122}\right),\label{eq:H0k}
 \end{align}
 \begin{align}
\mc H^{\pi}(\mathbf k) & =-\frac{1}{2}\left[\cos(k_{x}+k_{y})\Lambda^{001}+\sin(k_{x}+k_{y})\Lambda^{002}\right]\nonumber \\
 & \quad+\frac{1}{2}\left[\cos(k_{x}+k_{y})\Lambda^{031}+\sin(k_{x}+k_{y})\Lambda^{032}\right]\nonumber \\
 & \quad\,-\frac{1}{2}\left[\cos(k_{x}+k_{y})\Lambda^{301}+\sin(k_{x}+k_{y})\Lambda^{302}\right]\nonumber \\
 & \quad\,-\frac{1}{2}\left[\cos(k_{x}+k_{y})\Lambda^{331}+\sin(k_{x}+k_{y})\Lambda^{332}\right]\nonumber \\
 & \quad+\sin(k_{z})\left(-\cos k_{x}\Lambda^{211}+\sin k_{x}\Lambda^{212}\right)\nonumber \\
 & \quad+\cos(k_{z})\left(-\cos k_{x}\Lambda^{221}+\sin k_{x}\Lambda^{222}\right)\nonumber \\
 & \quad\,-\sin(k_{z})\left(\sin k_{y}\Lambda^{311}+\cos k_{y}\Lambda^{312}\right)\nonumber \\
 & \quad+\cos(k_{z})\left(\sin k_{y}\Lambda^{321}+\cos k_{y}\Lambda^{322}\right).
\end{align}

\begin{figure}
\begin{centering}
\includegraphics[width=0.5\columnwidth]{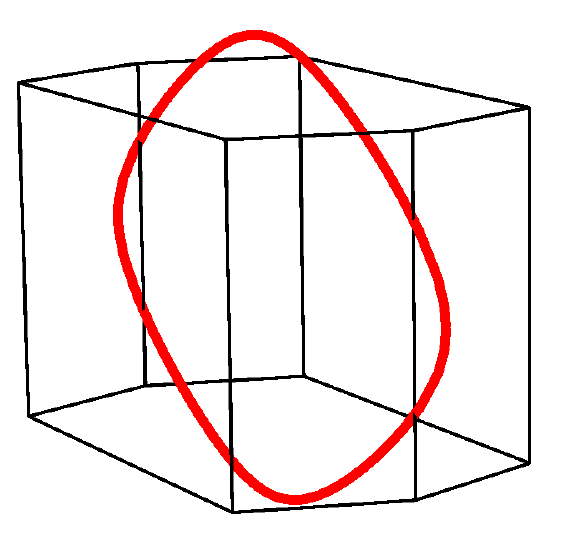}
\par\end{centering}
\caption{\label{fig:nodalline} (Color online) Nodal line (red) of the zero flux state. The thin black lines represent the edges of the first Brillouin zone of the base-centered orthorrombic lattice. }
\end{figure}

Figure \ref{fig:nodalline} shows the nodal line of the zero-flux state. We note that the   nodal line occurs on the boundary of  the first Brillouin zone of the face-centered orthorrombic lattice in the four-sublattice representation of the hyperhoneycomb lattice, cf. Ref. \cite{Lee2014}.

\section{Spin-Orbital Operators in terms of SU(4) generators\label{sec:LFWT_operators}}

Here we present the 15 operators $s^{a}$, $\tau^{b}$
and $s^{a}\tau^{b}$ in terms of the SU(4) generators 
$S_{m}^{n}$ defined in Eq. (\ref{generators}): 
\be
s^{x}  =\frac{1}{2}\underset{m=1,3}{\sum}\left(S_{m}^{m+1}+S_{m+1}^{m}\right),
\ee
\be
s^{y} =\frac{1}{2i}\underset{m=1,3}{\sum}\left(S_{m}^{m+1}-S_{m+1}^{m}\right),
\ee
\be
s^{z} =\frac{1}{2}\underset{m=1,3}{\sum}\left(S_{m}^{m}-S_{m+1}^{m+1}\right),
\ee
\be
\tau^{x} =\frac{1}{2}\underset{n=1,2}{\sum}\left(S_{n}^{n+2}+S_{n+2}^{n}\right),
\ee
\be
\tau^{y} =\frac{1}{2i}\underset{n=1,2}{\sum}\left(S_{n}^{n+2}-S_{n+2}^{n}\right),
\ee
\be
\tau^{z} =\frac{1}{2}\underset{n=1,2}{\sum}\left(S_{n}^{n}-S_{n+2}^{n+2}\right),
\ee
\be
s^{x}\tau^{x} =\frac{1}{4}\left(S_{1}^{4}+S_{2}^{3}+h.c.\right),
\ee
\be
s^{x}\tau^{y} =\frac{1}{4i}\left(S_{1}^{4}+S_{2}^{3}\right)+h.c.,
\ee
\be
s^{x}\tau^{z} =\frac{1}{4}\left(S_{1}^{2}-S_{3}^{4}+h.c.\right),
\ee
\be
s^{y}\tau^{x} =\frac{1}{4i}\left(S_{1}^{4}-S_{2}^{3}\right)+h.c.,
\ee
\be
s^{y}\tau^{y} =\frac{1}{4}\left(-S_{1}^{4}+S_{2}^{3}+h.c.\right),
\ee
\be
s^{y}\tau^{z} =\frac{1}{4i}\left(S_{1}^{2}-S_{3}^{4}\right)+h.c.,
\ee
\be
s^{z}\tau^{x} =\frac{1}{4}\left(S_{1}^{3}-S_{2}^{4}+h.c.\right),
\ee
\be
s^{z}\tau^{y} =\frac{1}{4i}\left(S_{1}^{3}-S_{2}^{4}\right)+h.c.,
\ee
\be
s^{z}\tau^{z} =\frac{1}{4}\left(S_{1}^{1}-S_{2}^{2}-S_{3}^{3}+S_{4}^{4}\right).
\ee

\bibliographystyle{apsrev4-1}
\bibliography{article2}

\end{document}